\documentclass[%
reprint,
superscriptaddress,
 amsmath,amssymb,
 aps,
prd,
]{revtex4-2}

\usepackage{graphicx}
\usepackage{dcolumn}
\usepackage{bm}

\usepackage{xcolor} 

\begin{document}


\title{Algorithm for TDI numerical simulation and sensitivity investigation}

\author{Gang Wang}
\email[Gang Wang: ]{gwang@shao.ac.cn, gwanggw@gmail.com}
\affiliation{Shanghai Astronomical Observatory, Chinese Academy of Sciences, Shanghai 200030, China}

\author{Wei-Tou Ni}
\email[Wei-Tou Ni: ]{weitou@gmail.com}
\affiliation{National Astronomical Observatories, Chinese Academy of Sciences, Beijing, 100012, China}
\affiliation{State Key Laboratory of Magnetic Resonance and Atomic and Molecular Physics, Innovation Academy for Precision Measurement Science and Technology (APM), Chinese Academy of Sciences, Wuhan 430071, China}
\affiliation{Department of Physics, National Tsing Hua University, Hsinchu, Taiwan, 30013, ROC}

\author{Wen-Biao Han}
\email[Wen-Biao Han: ]{wbhan@shao.ac.cn}
\affiliation{Shanghai Astronomical Observatory, Chinese Academy of Sciences, Shanghai 200030, China}
\affiliation{School of Astronomy and Space Science, University of Chinese Academy of Sciences, Beijing 100049, China}
\affiliation{School of Fundamental Physics and Mathematical Sciences, Hangzhou Institute for Advanced Study, UCAS, Hangzhou 310024, China}
\affiliation{International Centre for Theoretical Physics Asia-Pacific, Beijing/Hangzhou, China}
\affiliation{Key Laboratory for Research in Galaxies and Cosmology, Shanghai Astronomical Observatory, Shanghai 200030, China}

\author{Cong-Feng Qiao}
\email[Cong-Feng Qiao: ]{qiaocf@ucas.ac.cn}
\affiliation{School of Physics, University of Chinese Academy of Sciences, Beijing 100049, China}
\affiliation{CAS Center for Excellence in Particle Physics, Beijing 100049, China}

\date{\today}

\begin{abstract}

We introduce a generic algorithm to determine the time delays and spacecraft (S/C) positions to compose any time-delay interferometry (TDI) channel in the dynamical case and evaluate its sensitivity by using a full numerical method. We select 11 second-generation TDI channels constructed from four approaches and investigate their gravitational wave responses, noise levels, and averaged sensitivities under a numerical LISA orbit. The sensitivities of selected channels are various especially for frequencies lower than 20 mHz. The optimal channel A$_2$ (or equivalently E$_2$) combined from second-generation Michelson TDI channels (X$_1$, X$_2$, and X$_3$) achieves the best sensitivity among the channels, while the Sagnac $\alpha_1$ channel shows the worse sensitivity. Multiple channels show better sensitivities at some characteristic frequencies compared to the fiducial X$_1$ channel. The joint $\mathrm{A_2+E_2+T_2}$ observation not only enhances the sensitivity of the X$_1$ channel by a factor of $\sqrt{2}$ to 2 but also improves the capacity of sky coverage.

\end{abstract}

\keywords{Gravitational Wave, Time-Delay Interferometry, LISA, TAIJI }

\maketitle


\section{Introduction}

Advanced LIGO and Advanced Virgo have observed a score of gravitational wave (GW) signals since the first detection of binary black hole (BH) coalescence--GW150914 \cite[and references therein]{Abbott:2016blz,TheLIGOScientific:2016pea,TheLIGOScientific:2017qsa,LIGOScientific:2018mvr,Nitz:2019hdf,Abbott:2020uma,LIGOScientific:2020stg,Abbott:2020khf,Abbott:2020niy}. Most of the detections were identified as binary BH system, and two events were recognized as binary neutron star coalescence \cite{TheLIGOScientific:2017qsa,Abbott:2020uma}. The recently announced detection, GW190521, was inferred as intermediate mass BH coalescence \cite{Abbott:2020tfl}.

LISA is scheduled to be launched around the 2030s and is targeting to detect the GW in frequency 0.1 mHz to 100 mHz. By employing the drag-free technology, three spacecraft (S/C) follow their respective geodesics to form a triangular laser interferometer with an arm length of $2.5 \times 10^6$ km. If traditional laser metrology is utilized for the long and unequal interferometric arms, laser frequency noise will be too overwhelming to detect the GW signals. To achieve the targeting sensitivity, time-delay interferometry (TDI) is proposed for the LISA to suppress the laser frequency noise. 
In the previous literature, TDI had been well studied and demonstrated for the first-generation \cite[and references therein]{1999ApJ...527..814A,2000PhRvD..62d2002E,2001CQGra..18.4059A,Larson:2002xr,Dhurandhar:2002zcl,2003PhRvD..67l2003T,Vallisneri:2004bn,2008PhRvD..77b3002P,Tinto:2020fcc} and the second-generation \cite[and references therein]{Shaddock:2003dj,Cornish:2003tz,Tinto:2003vj,Dhurandhar:2010pd,Tinto:2018kij,2019PhRvD..99h4023B,2020arXiv200111221M,Vallisneri:2020otf}. The first-generation TDI configurations are designed to cancel the laser noise for a static constellation. The second-generation TDI is proposed to suppress laser frequency fluctuations in the case of time-varying arm lengths up to the first order derivative and solving a more realistic situation.

With the implementation of TDI, the secondary noises and GW signals could be accumulated or canceled with the paths combination. The acceleration noise and optical path noise are supposed to be the leading noises in the second-generation TDI. The cancellations/suppressions of other secondary noises are in an active study stage. For instance, the clock jitter noise could be reduced using new measurement combinations \cite{Otto:2012dk,Otto:2015,Tinto:2018kij,Hartwig:2020tdu}, and the tilt-to-length noise could be resolved by a new optical designs \cite{Chwalla:2016bzk,Trobs:2017msu}. The GW response in the TDI channels could also be suppressed/enhanced by the path combination. To investigate the noise level and GW response in TDI channels, multiple simulators have been developed for LISA mission with the different focuses \cite{2003PhRvD..67b2001C,2004PhRvD..69h2003R,Vallisneri:2004bn,2008PhRvD..77b3002P,2019PhRvD..99h4023B}. Considering the complexity of the TDI calculation, most of the investigations implemented the analytical or semi-analytical algorithms for the calculations.

We have developed a numerical algorithm to calculate the path mismatches in TDI for LISA-like missions and ASTROD-GW concept since 2011 \cite{Ni:2013,Wang:2011,Wang:2014aea,Wang:2012ce,Wang:2012te,Dhurandhar:2011ik,Wang:2014cla,Wang:2017aqq,Wang:2019ipi}. In previous work \cite{Wang:2020fwa}, by using a set of numerical orbit, we investigated the GW responses, noise levels and  sensitivities of the first-generation TDI channels for LISA and TAIJI. We adopted a semi-analytical approach to evaluate the GW response and secondary noises in an instantaneous static TDI configuration, and implemented the numerical method to calculate the laser frequency noise raised by the path mismatches in a dynamic TDI configuration. For the second-generation TDI, their configurations could be flexibly constructed from different approaches \cite[and references therein]{Shaddock:2003dj,Tinto:2003vj,Vallisneri:2005ji,Dhurandhar:2010pd,2020arXiv200111221M}. Therefore, due to the versatility and flexibility of path combinations, the complexity of analytical evaluation would significantly increase to adapt to various TDI channels. Furthermore, compared to the first-generation TDI, the second-generation TDI would involve more links in a wider time span, the static approximation would increase the inaccuracy of the calculations.

To investigate TDI performances in the realistic dynamical case, by using numerical mission orbit and an ephemeris framework, we start from our previous algorithm which can determine the S/C positions and laser propagation time between S/C incorporating relativistic time delays during TDI. And we develop new modules in this work to thoroughly evaluate the GW response, noise level, and sensitivity for any TDI observable by using a numerical algorithm. As a preparation for algorithm implementation, a S/C layout-time delay diagram is employed to illustrate the paths of a TDI channel and streamline the calculations procedures.
We select 11 second-generation TDI channels constructed from four approaches, and investigate their yearly averaged sensitivities. The investigations show that mismatches of laser paths in selected channels are sufficiently decreased and could make the laser frequency noise well below the secondary noises.
The optimal channels A$_2$/E$_2$ combined from second-generation Michelson TDI channels (X$_1$, X$_2$, and X$_3$) achieves the best sensitivity in selected channels for frequencies lower than 50 mHz, while the Sagnac $\alpha_1$ channel shows the worse sensitivity. Multiple TDI channels could have better sensitivities at some characteristic frequencies compared to the fiducial X$_1$ channels. The sensitivities of Michelson-type TDI channels would be identical considering the noise level increase/decrease with the GW response increase/decrease.

This paper is organized as follows. 
In Sec. \ref{sec:algorithm}, we introduce the numerical algorithm to determine the time delay and S/C positions in TDI, and the selected TDI channels constructed from different approaches. 
In Sec. \ref{sec:response}, we present the TDI response calculation by using the obtained time delay and S/C positions, and analyze the yearly averaged responses for the selected channels.
In Sec. \ref{sec:noise}, we examine laser frequency noise due to the path mismatch in TDI channels, and numerically evaluate the noise levels generated from the acceleration noise and optical path noise.
In Sec. \ref{sec:sensitivity}, we synthesize yearly averaged sensitivities of each TDI channel, and compare their sensitivities with the fiducial first-generation Michelson X channel. 
We give our conclusions in Sec. \ref{sec:conclusions}.
(We set $G=c=1$ in this work except specified in the equations.)

\section{Time delay determination and TDI channel selections} \label{sec:algorithm}

\subsection{Algorithm for TDI diagram} \label{subsec:algorithm}

The purpose of TDI is to construct the equivalent equal arm interferometer by combining multiple arm links in sequence and cancel laser frequency noise. The path matching of a TDI configuration depends on the arm lengths and relative motions between the S/C. A numerical LISA orbit is utilized in this investigation as shown in Fig. 2 of \cite{Wang:2020a}.
The orbit is achieved based on the LISA 2017 requirements that the amplitudes of relative velocities between S/C are less than 5 m/s for $2.5 \times 10^6$ km arm length \cite{2017arXiv170200786A}.
The original LISA optical design is to equip two optical benches on each S/C and collects two measurements on each optical bench \cite[and references therein]{2000PhRvD..62d2002E,Vallisneri:2004bn,2008PhRvD..77b3002P}. A newly designed configuration has been proposed that three/four measurements are gathered on each optical bench to remove some secondary noises more effectively \cite[and references therein]{Otto:2012dk,Otto:2015,Tinto:2018kij,2019PhRvD..99h4023B}. And we employ the newly designed measurements on each optical bench in this work.

The expressions of the first-generation TDI channels have been formulated in previous works \cite{1999ApJ...527..814A,2000PhRvD..62d2002E,2001CQGra..18.4059A,Larson:2002xr,2003PhRvD..67l2003T,Tinto:2003vj,Shaddock:2003dj,Vallisneri:2004bn,2008PhRvD..77b3002P,Dhurandhar:2010pd,Tinto:2020fcc}, for instance, the expression of measurements in the Michelson-X channel could be simplified as
\begin{equation} \label{eq:X_measurement}
\begin{aligned}
{\rm X} =& [ \eta_{31} + \mathcal{D}_{31} \eta_{13}  + \mathcal{D}_{13} \mathcal{D}_{31} \eta_{21} + \mathcal{D}_{21} \mathcal{D}_{13} \mathcal{D}_{31} \eta_{12}  ] \\
 & -  [ \eta_{21} + \mathcal{D}_{21} \eta_{12} + \mathcal{D}_{12} \mathcal{D}_{21} \eta_{31} + \mathcal{D}_{31} \mathcal{D}_{12} \mathcal{D}_{21} \eta_{13}  ], \\
\end{aligned}
\end{equation}
by implementing the combined observables $\eta_{ji}$ for S/C$j$ to S/C$i$ ($j \rightarrow i$) as defined in \citep{Otto:2012dk,Otto:2015,Tinto:2018kij},
\begin{equation} \label{eq:eta}
\begin{aligned}
  \eta_{ji} &= s_{ji} + \frac{1}{2} \left[ \tau_{ij} - \varepsilon_{ij} + \mathcal{D}_{ji} ( 2 \tau_{ji} - \varepsilon_{ji} - \tau_{jk} ) \right] \\
  & \quad \mathrm{for} \  (2 \rightarrow 1), (3 \rightarrow 2) \ \mathrm{and} \ (1 \rightarrow 3), \\ 
  \eta_{ji} &= s_{ji} + \frac{1}{2} \left[ \tau_{ij} - \varepsilon_{ij}  + \mathcal{D}_{ji} ( \tau_{ji} - \varepsilon_{ji} ) + \tau_{ik} -  \tau_{ij} \right] \\
   & \quad \mathrm{for} \  (1 \rightarrow 2), (2 \rightarrow 3)\ \mathrm{and}\ (3 \rightarrow 1),
\end{aligned}
\end{equation}
where $s_{ji}$, $\varepsilon_{ij}$ and $\tau_{ij}$ are described in Appendix \ref{sec:appendix_observables}, $\mathcal{D}_{ij}$ is a time-delay operators and act on a measurement $y(t)$ by 
\begin{equation}
\begin{aligned}
 \mathcal{D}_{ij} y(t) &= y(t - L_{ij}(t) ), \\
 \mathcal{D}_{mn}\mathcal{D}_{ij} y(t) &= y( t - L_{ij}(t) - L_{mn}( t - L_{ij}(t) ) ), \\
 & \cdots  \cdots
\end{aligned}
\end{equation} 

\begin{figure}[htb]
\includegraphics[width=0.238\textwidth]{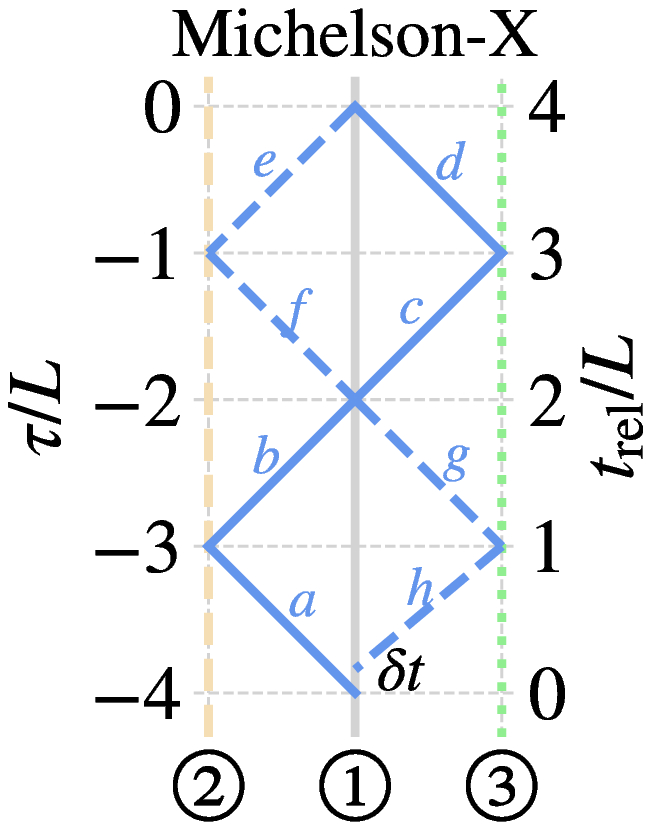} 
\includegraphics[width=0.238\textwidth]{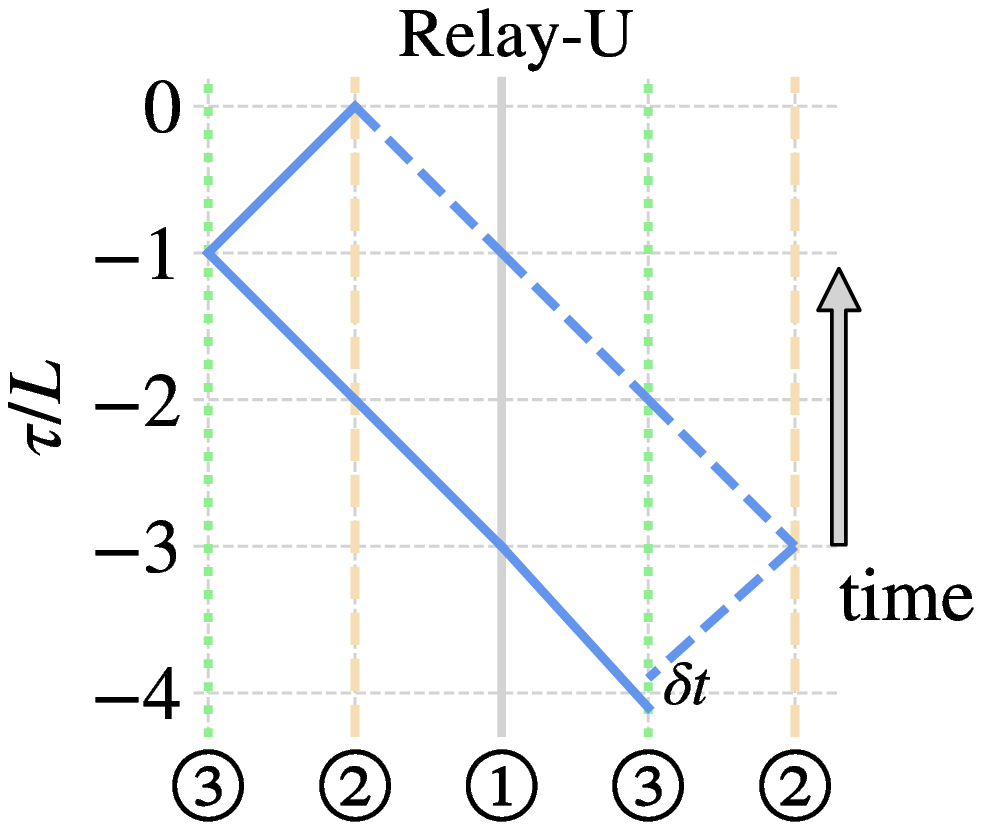} 
\includegraphics[width=0.238\textwidth]{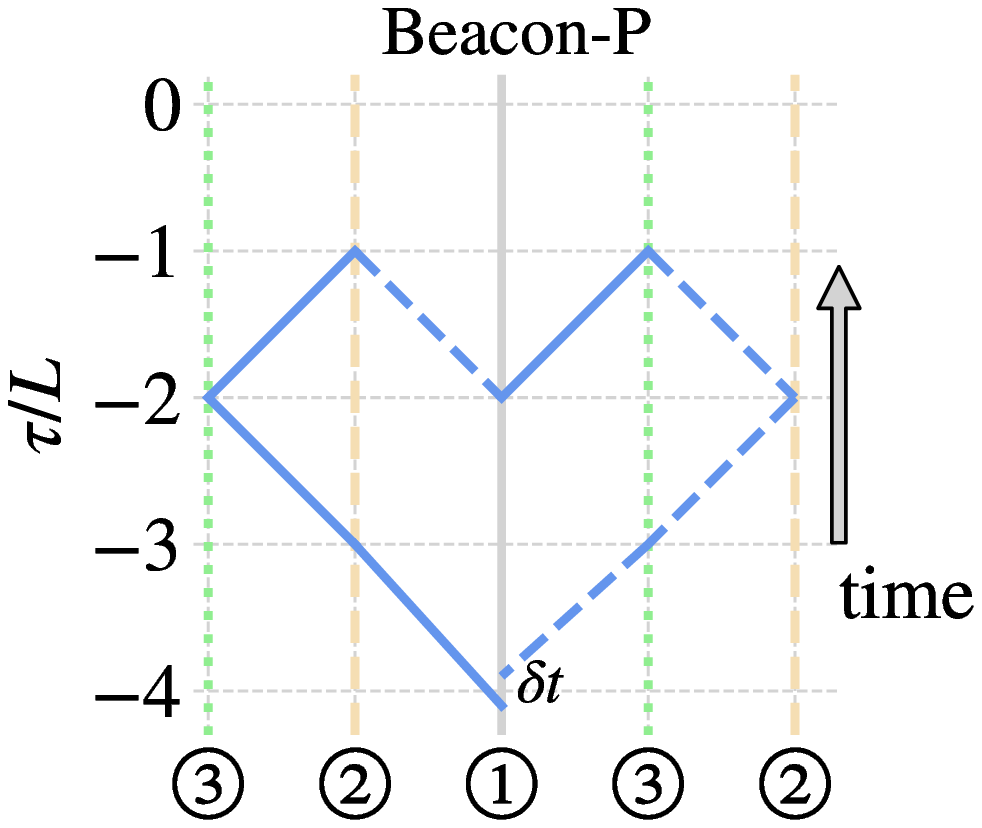}
\includegraphics[width=0.238\textwidth]{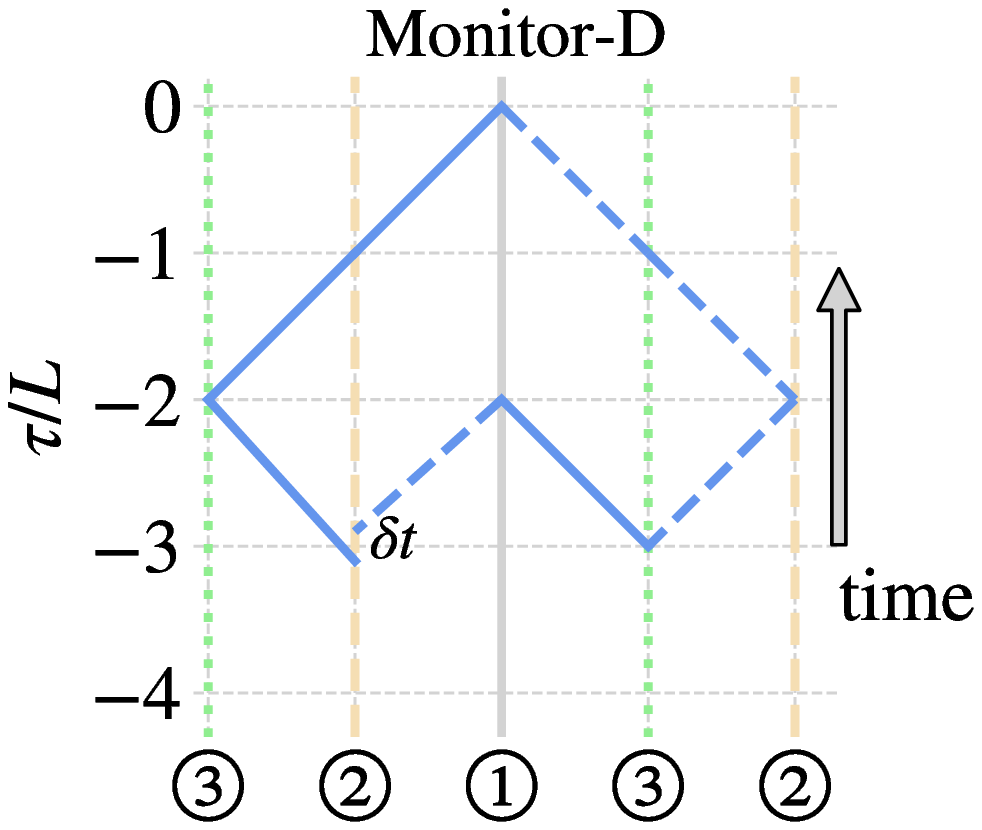}
\caption{\label{fig:1st_TDI_diagram} The S/C layout-time delay diagrams for the first-generation TDI channels Michelson-X, Relay-U, Beacon-P, and Monitor-D. The vertical lines indicate the trajectories of S/C on time direction (\textcircled{$i$} indicates S/C$i$, $i = 1, 2, 3$), and the ticks on each left y-axis show the value of time delay with respect to the TDI ending time $\tau=0$. The $t_\mathrm{rel}$ is the time with respect to the starting time $t_0$ at starting S/C of TDI. To avoid the cross caused by TDI paths at non-integer delay time and show the paths tidy, extra trajectory lines are plotted for S/C2 (dotted green lines) and S/C3 (dashed orange lines). The blue lines show the paths of TDI channels, the solid line and dashed line indicate two groups of interfered laser beams.}
\end{figure}

To visualize the paths of TDI and streamline the calculations, we developed a S/C layout-time delay diagrams for TDI configurations \cite{Wang:2011}, \citet{2020arXiv200111221M} developed their similar space-time diagram in recent work. The diagrams for four first-generation TDI channels are shown in Fig. \ref{fig:1st_TDI_diagram}. The x-axis shows the spatial separation between S/C, and the y-axis shows the time direction. Each vertical line indicates the trajectory of one S/C with time (\textcircled{$i$} indicates S/C$i$, $i = 1, 2, 3$), and the ticks on each left y-axis show the value of time delay with respect to the TDI ending time $\tau=0$. We clarify that $t_\mathrm{rel}$ is the time with respect to the starting time $t_0$ at starting S/C of TDI.
To avoid paths cross at non-integer delay time and show the paths tidy, the extra trajectory lines are plotted for S/C2 (dotted green lines) and S/C3 (dashed orange lines). The blue lines show the paths of the TDI channels, the solid line and dashed line indicate two groups of interfered laser beams, as well as the signs in the TDI expression.

We select the Michelson-X to specify the procedures of diagram implementation as shown in the upper left plot of Fig. \ref{fig:1st_TDI_diagram} and Table \ref{tab:TDI_X}. The calculation starts from the mission time $t_0=0$ at point $\xi( \textcircled{1}; t_{\rm{rel}} = 0; \tau \simeq -4L ) $ toward the $\xi( \textcircled{2}; t_{\rm{rel}} =  L_{12}; \tau \simeq -3L ) $ (the link $a$, corresponding to the last term in the first row of Eq. \eqref{eq:X_measurement}), and determine the relative time $ t_{\rm{rel}} =  8.3474947$ s with respect to $t_0$ and register the position of \textcircled{2} at the time $t_0 + t_{\rm{rel}} $. Along the link $b$, the second step starts from $\xi( \textcircled{2}; t_{\rm{rel}} =  8.3474947 \mathrm{\ s}; \tau \simeq -3L )$  to $\xi( \textcircled{1}; t_{\rm{rel}} =  L_{12} + L_{21}; \tau \simeq -2L )$, and update the relative time $t_{\rm{rel}} = 16.6934247 $ s and register the position of \textcircled{1} at updated $t_0 + t_{\rm{rel}} $. The following steps are implemented based on alphabetical order from the link $c$ until final link $h$ by using the same method. For the links with backward time direction, a minus sign will be implemented which corresponds to the terms with minus signs in Eq. \eqref{eq:X_measurement}.
The process of each step could be expressed as
\begin{equation}
 l\mathrm{th\ link}:  \xi(  {\bm{r}}_{i,l-1} ; t_{\mathrm{rel}, l-1} ;  \tau_{i,l-1} )  \mapsto  \xi( {\bm{r}}_{j,l} ; t_{\mathrm{rel}, l} ; \tau_{j,l} ). 
\end{equation}

After all steps are implemented, the final ending S/C will be the initial sending S/C. The relative time $t_{\rm rel}$ would differ from starting $t_{\rm rel} = 0$ due to the relative motion between S/C during TDI which is the path mismatch $\delta t$. In previous works \cite{Wang:2011,Wang:2014aea,Wang:2012ce,Wang:2012te,Dhurandhar:2011ik,Wang:2014cla,Wang:2017aqq,Wang:2019ipi}, we implemented this algorithm to calculate the path mismatch for laser frequency noise suppression and verify the feasibility of TDI configurations. Another purpose of the calculation is to determine time delay in each link and the S/C positions, since the response to a GW signal will depend on the time delay factors and instantaneous positions of S/C, and the noises in TDI channels are related to the time delay.

For a TDI channel, the time at the top vertex point is defined as $t_{m} = t_0 + \max(t_{\rm rel})$ and time delay is set to be $ \tau = 0$ except for Beacon-P configuration, and then the time delay at each step is calculated by $t_{\rm rel} - \max(t_{\rm rel})$. The results of each step for the Michelson-X channel at starting mission time are shown in Table \ref{tab:TDI_X}.
\begin{table}[tbh]
\caption{\label{tab:TDI_X}
The results of the first-generation Michelson-X channel calculation in each step at staring mission time $t_0 = 0$. The delay time is determined from the relative time $t_\mathrm{rel}$ by $\tau = t_{\rm rel} - \max(t_{\rm rel})$. (Only the first 7 decimals are present for time factors).}
\begin{ruledtabular}
\begin{tabular}{cccccc}
step &  mission time & relative time &  delay time  & S/C  & Position \\
$l$ & $t_0$ & $t_{\rm rel}$ (s) & $\tau$ (s)  & & (SSB)\footnotemark[1]  \\
\hline
0 & 0 & 0 & -33.4510480	& 1 & ${\bm{r}}_1 $ \\
1 & 0 & 8.3474947 & -25.1035533 &	2	& ${\bm{r}}_2 $ \\
2 & 0 & 16.6934247 & -16.7576233	& 1 & ${\bm{r}}_1 $ \\
3 & 0 & 25.0728702 & -8.3781778 & 3 & ${\bm{r}}_3 $ \\
4 & 0 & 33.4510480	& 0	& 1 & ${\bm{r}}_1 $ \\
5 & 0 & 25.1051181	& -8.3459299	& 2	& ${\bm{r}}_2 $ \\
6 & 0 & 16.7576236 & -16.6934244	& 1 & ${\bm{r}}_1 $ \\
7 & 0 & 8.3794460 & -25.0716020 &3 & ${\bm{r}}_3 $ \\
8 & 0 & 6.3697e-7 & -33.4510474 & 1 & ${\bm{r}}_1 $ \\
\end{tabular}
\end{ruledtabular}
\footnotetext[1]{the positions in the solar-system barycentric coordinates.}
\end{table}

The combined measurement of one TDI channel is the sum of each step,
\begin{equation}  \label{eq:TDI_combination}
 \mathrm{TDI} = \sum^{n}_{l = 1} \mathrm{sgn}( \tau_{l} - \tau_{l-1} ) \eta_{ij,l}(t_m + \tau_{j,l} ),
\end{equation}
with sign function
\begin{equation}
\begin{aligned}
& \mathrm{sgn}( \tau_{l} - \tau_{l-1} )  = \left\{
  \begin{aligned}
  1 & \quad \mathrm{if} \quad  \tau_{l} - \tau_{l-1} > 0, \\ 
  -1 & \quad \mathrm{if} \quad  \tau_{l} - \tau_{l-1} < 0.
\end{aligned}  \right.
\end{aligned}
\end{equation}

\subsection{Time delay calculation}

The time delay between the laser beam sender and receiver is essential for TDI calculation as aforementioned. Due to the gravitational field of celestial bodies, there will be extra relativistic time delay during the light propagation besides the delay from coordinate distance, and its leading order is required to be considered during the calculation \cite{Ashby:2008lea}. The time delay from the sending time $T^s$ at $\bm{r}_1$ to the receiving time $T^r$ at $\bm{r}_2$ is calculated by \cite{Shapiro:1964uw,Kopeikin:2008xv},
\begin{equation} \label{eq:L_ij}
 T^{r} - T^{s} =  \frac{R}{c} + \Delta T_{\text{PN}},
\end{equation}
where $R$ is the coordinate distance between the sender and receiver S/C, $c$ is speed of light, and $\Delta T_{\text{PN}}$ is the relativistic time delay caused by the gravitational field, 
   \begin{equation}
   \begin{split}
   \Delta T_{\text{PN}} & = \frac{2GM}{c^3} \ln \left( \frac{R_1+R_2+R}{R_1+R_2-R} \right) \\
    + &  \frac{G^2 M^2}{c^5} \frac{R}{R_1 R_2} \left[ \frac{15}{4} \frac{\arccos({\bf{N}}_1 \cdot {\bf{N}}_2 ) }
    {|{\bf{N}}_1 \times {\bf{N}}_2|} -\frac{4}{1+ \bf{N}_1 \cdot \bf{N}_2 } \right],
    \end{split}
   \end{equation}
where $G$ is gravitational constant, $M$ is the gravitational body, $\bf{N}_1$ and $\bf{N}_2$ are the respective unit vector from the gravitating body to the sender and receiver, and $R_1$ and $R_2$ is the radial distances of sender and receiver from gravitating body. The leading order relativistic time delay caused by gravitational field of the Sun is included in our current calculation, the effects from other planets should be orders lower than the Sun's for LISA mission. 

On the other side, due to the relative motion between S/C, the displacement of receiver during the light propagation is also considered. The receiving time is determined by using iteration in numerical calculation,
 \begin{equation}
 \begin{aligned}
      T^r_0 &= T^s_0 + T_1 + T_2 + T_3 +...  \\
       T_1      &= \frac{| \bm{r}_r (T^s_0)-\bm{r}_s (T^s_0) |}{c} + \Delta T_{1,\text{PN}}  \\
      T_1 + T_2  &= \frac{| \bm{r}_r (T^s_0 +T_1)-\bm{r}_s (T^s_0) |}{c} + \Delta T_{2,\text{PN}}  \\
     T_1 + T_2 +T_3 &=\frac{| \bm{r}_r (T^s_0+T_1+T_2)-\bm{r}_s(T^s_0) |}{c} + \Delta T_{3,\text{PN}}  \\
   & ......
\end{aligned}
\end{equation}
During the iteration calculation, the Chebyshev polynomial interpolation is utilized to precisely obtain the position of S/C at any moment \cite{Newhall:1989CeMec,Li&Tian:2004}.

\subsection{Selection of TDI channels}

For the first-generation TDI, there are five recognized configurations which are Sagnac, Michelson, Relay, Beacon, and Monitor. However, the configurations of the second-generation TDI are more flexible and could be constructed from different approaches \cite[and references therein]{Shaddock:2003dj,Tinto:2003vj,Vallisneri:2005ji,Dhurandhar:2010pd,2020arXiv200111221M}. In this work, we select 11 typical TDI channels derived from four methods as follows.
\begin{itemize}
\item 
the first group of the second-generation TDI channels is derived from two same first-generation TDI channels, and the channels are constructed from two same first-generation channels with a relative time shift. The expressions could be described as,
\begin{equation} \label{eq:regular_TDI}
\begin{aligned}
 {\rm X_1}(t) & \approx  {\rm X}( t - 4 L ) - {\rm X}(t) ,  \\
  { \alpha_1}(t) & \approx  \alpha( t - 3 L ) - \alpha(t) ,  \\
   {\rm U_1}(t) & \approx {\rm U}( t - 2 L ) - {\rm U}(t) , \\
   {\rm P_1}(t) & \approx {\rm P}( t - L ) - {\rm P}(t + L),  \\
   {\rm D_1}(t) & \approx {\rm D}( t - 2 L ) - {\rm D}(t). 
\end{aligned}
\end{equation}
The subscript $1$ indicates the first channel of second-generation TDI combining from its first-generation family.
By using four first-generation TDI channels from each configuration shown in Fig. \ref{fig:1st_TDI_diagram}, their corresponding second-generation diagrams are shown in Fig. \ref{fig:2nd_TDI_diagram}. The approximation is used to emphasize that time delay between two TDI channels is not exactly equal to the integer times of arm length in numerical calculation. Similar to the first-generation, as we can expect, the performances of P/P$_1$ and D/D$_1$ will be identical. And we will choose P/P$_1$ to represent these two configurations in the following investigations.
\begin{figure}[thb]
\includegraphics[width=0.238\textwidth]{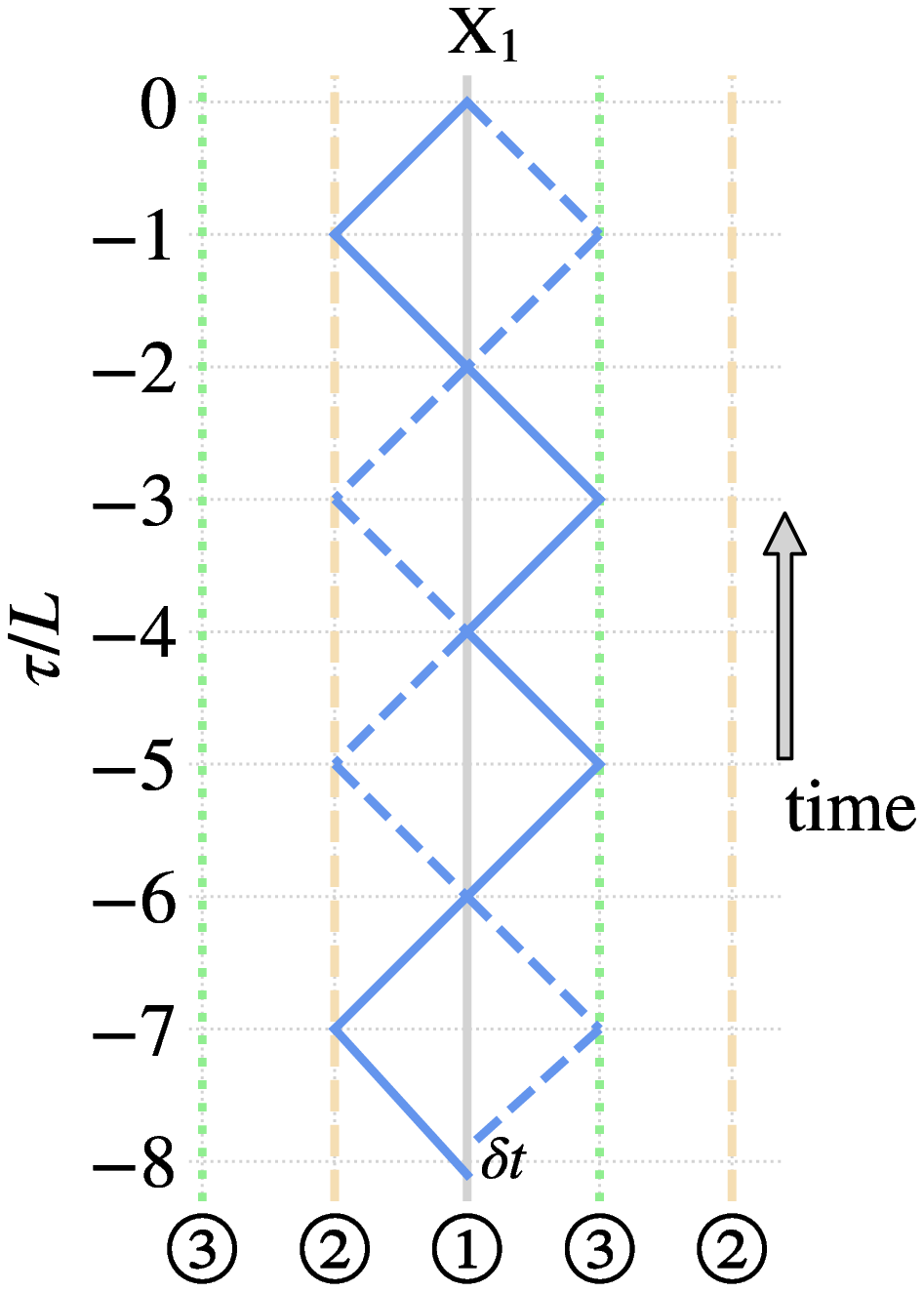} 
\includegraphics[width=0.238\textwidth]{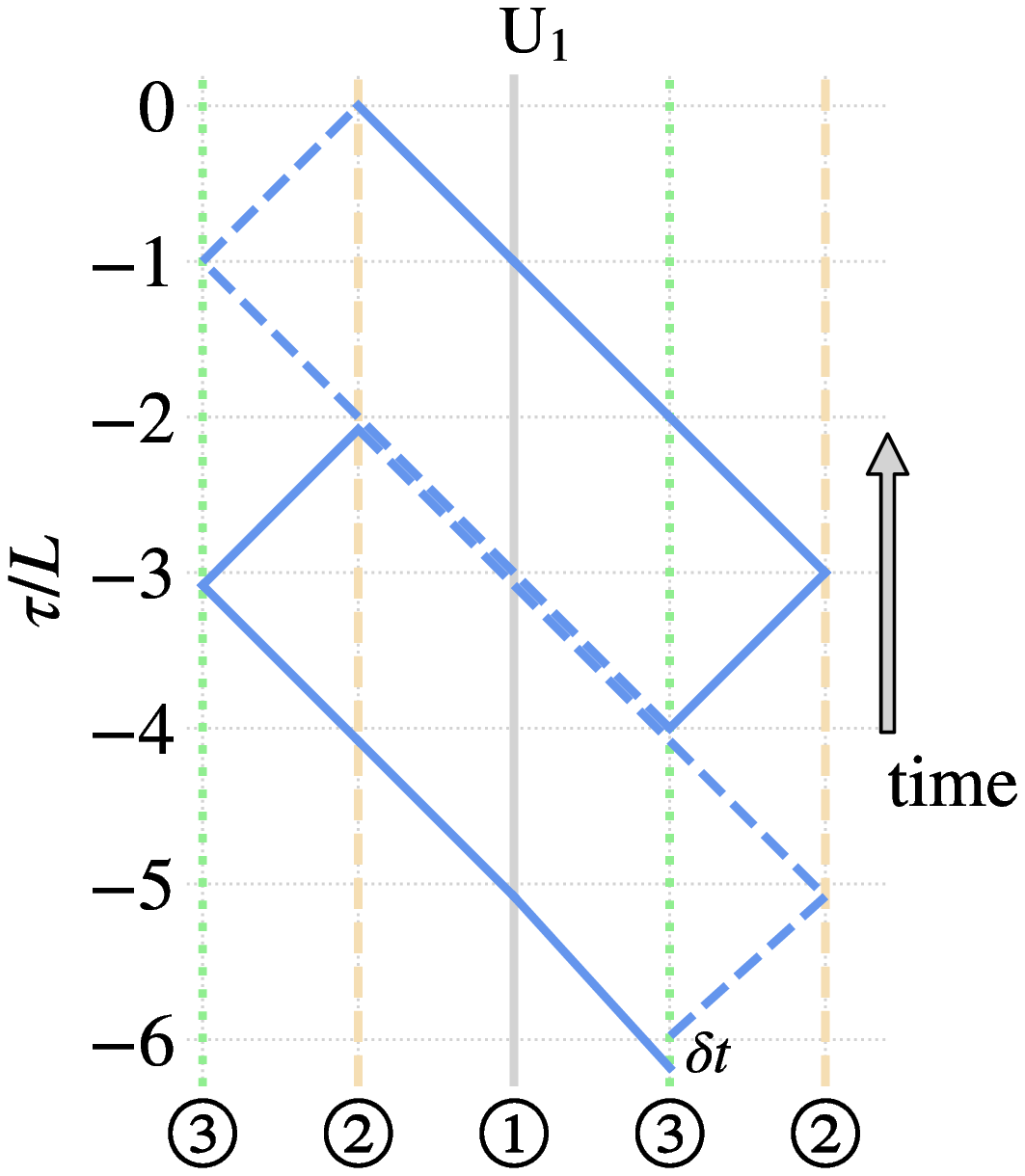} 
\includegraphics[width=0.238\textwidth]{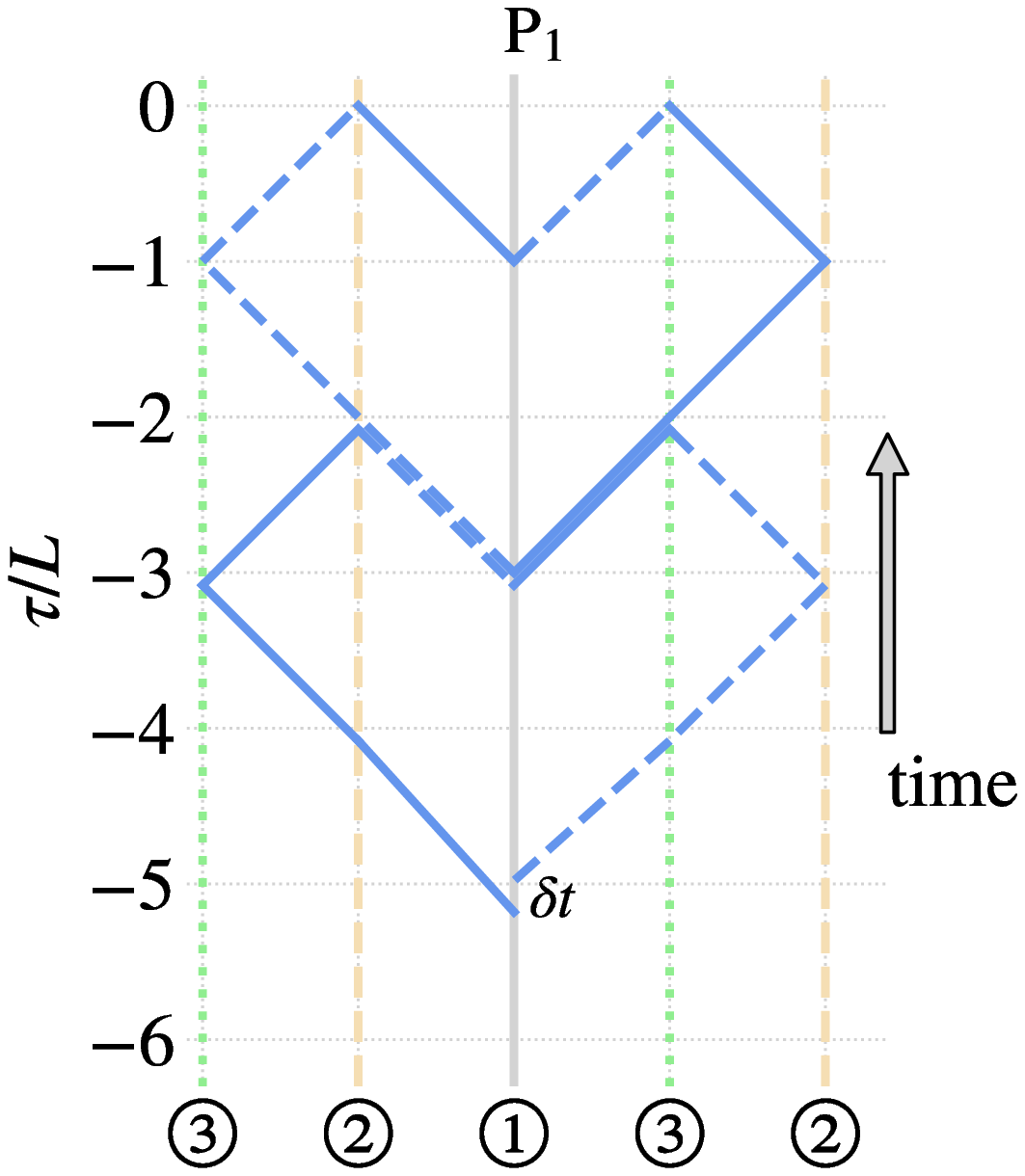}
\includegraphics[width=0.238\textwidth]{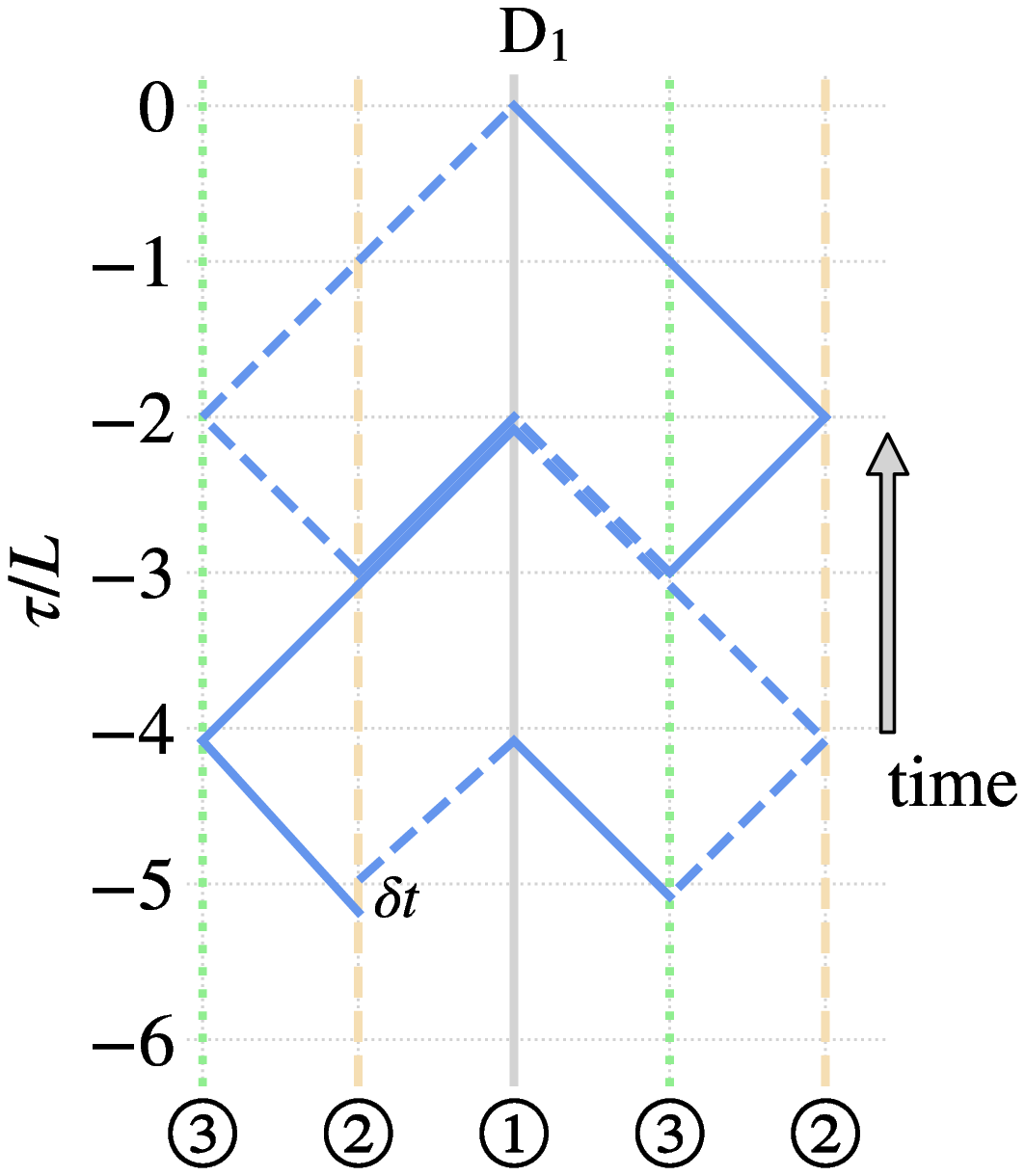}
\caption{\label{fig:2nd_TDI_diagram} The diagrams of the second-generation TDI channels Michelson-X$_1$, Relay-U$_1$, Beacon-P$_1$ and Monitor-D$_1$ constructed respectively from two same first-generation channels with a relative time shift.}
\end{figure}

\item
the second group of TDI channels is the optimal channels combined from three channels of one configuration. Similar to the optimal TDI channels, (A, E, and T), generated from three first-generation Michelson channels (X, Y, and Z) \cite{Prince:2002hp,Vallisneri:2007xa}, three optimal channels can be constructed from the second-generation Michelson channels ($\mathrm{X}_1$, $\mathrm{X}_2$, and $\mathrm{X}_3$, the $\mathrm{X}_2$ and $\mathrm{X}_3$ channels are obtained by cyclical permutation of the spacecraft indices from $\mathrm{X}_1$) by using the corresponding liner combinations,
\begin{equation} \label{eq:optimalTDI}
\begin{aligned}
 {\rm A_2} &=  \frac{ {\rm X_3} - {\rm X_1} }{\sqrt{2}} , \\
 {\rm E_2} &= \frac{ {\rm X_1} - 2 {\rm X_2} + {\rm X_3} }{\sqrt{6}} , \\
 {\rm T_2} &= \frac{ {\rm X_1} + {\rm X_2} + {\rm X_3} }{\sqrt{3}}.
\end{aligned}
\end{equation}
The A$_2$ and E$_2$ are also expected to have equivalent performance, and we choose the A$_2$ channel to represent their results.

\item 
the third group is extended TDI channels from two different first-generation channels. 
A second-generation TDI channel could be formed by combining one first-generation TDI channel with its time flipped one. The Monitor-D channel could be recognized as a time flipped Beacon-P as shown in Fig. \ref{fig:1st_TDI_diagram}, and their combination is named PD as shown in Fig. \ref{fig:2nd_TDI_diagram_others}. Moreover, the diagram could also indicate the reason for an identical performances of Beacon and Monitor as we have shown in \cite{Wang:2020fwa}. 
The $\overline{\mathrm{U}}$ denotes the flipped U channel, and the combined channel is named U$\overline{\mathrm{U}}$ as shown in Fig. \ref{fig:2nd_TDI_diagram_others}. Their expressions could be described as
\begin{equation}
\begin{aligned}
	{\rm PD}(t) & \approx  {\rm P}(t + L) + {\rm D}( t - 3 L ) , \\
   {\rm U\overline{U}}(t) & \approx {\rm U}( t - 4 L ) + \overline{{\rm U}}( t). \\
\end{aligned}
\end{equation}
Due to the symmetry of the Michelson and Sagnac configuration, the $\mathrm{X}_1$ and $\alpha_1$ essentially could also be obtained from this method. 
This approach could be extended to the various combination from any two or more channels from the 15 first-generation channels, and we only select these two channels in this work.
\begin{figure}[thb]
\includegraphics[width=0.238\textwidth]{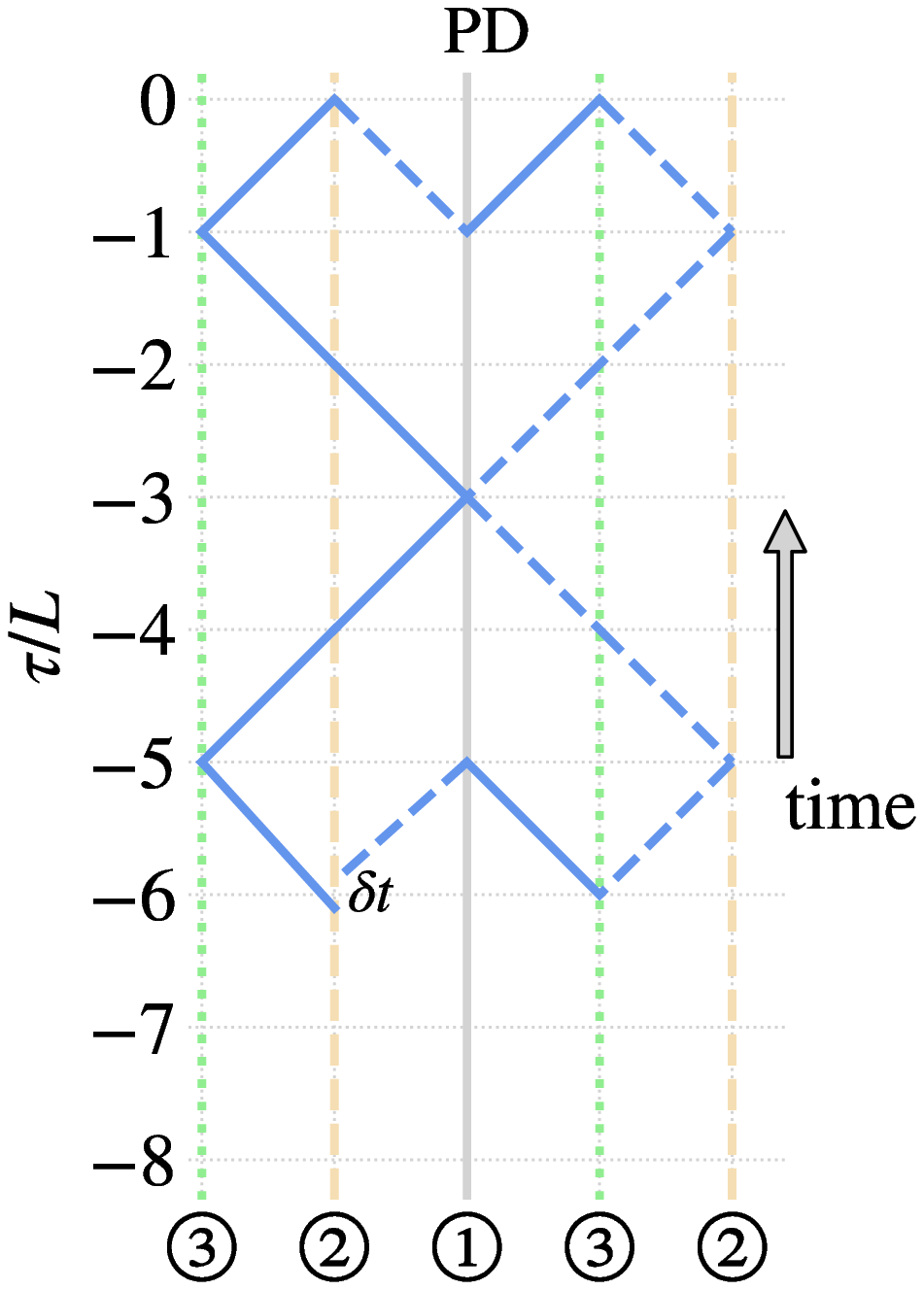} 
\includegraphics[width=0.238\textwidth]{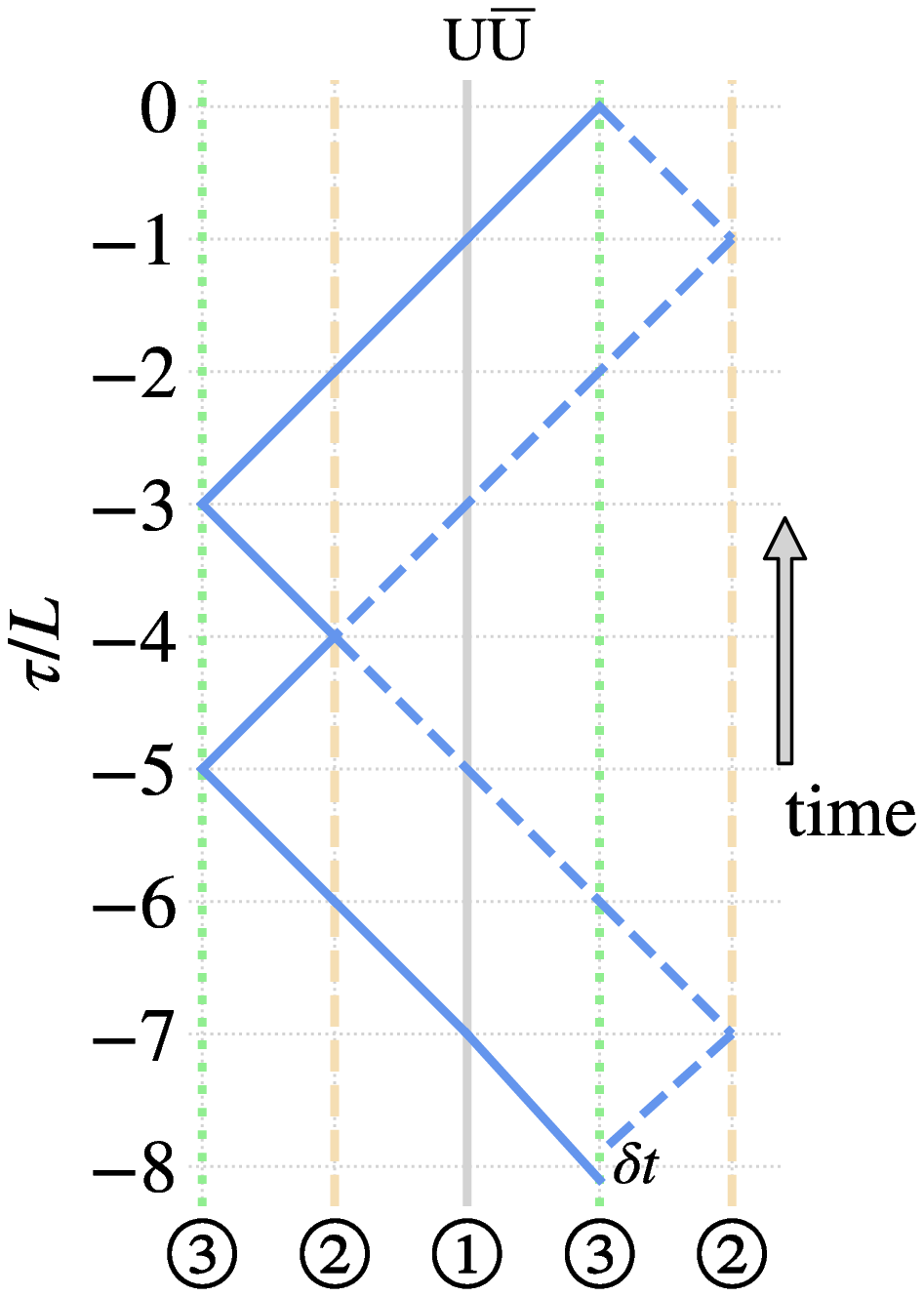} 
\caption{\label{fig:2nd_TDI_diagram_others} The diagrams of the  PD and U$\overline{\mathrm{U}}$ channels.}
\end{figure}

\item
the last group is the Michelson-type TDI channels proposed in \citet{Dhurandhar:2010pd} which only employ two interferometer arms. A bunch of TDI channels could be derived from two arms/four links by using this approach. By defining $a$ as the round trip along Arm3 ($a$: S/C1 $\rightarrow$ S/C2 $\rightarrow$ S/C1) and $b$ as the round trip along Arm2 ($b$: S/C1 $\rightarrow$ S/C3 $\rightarrow$ S/C1), the $\mathrm{X}_{aabb}$ channel is selected to be investigated and expressed as
\begin{equation} \label{eq:X_aabb}
 \mathrm{X}_{aabb} = [aabb, bbaa] \equiv aabbbbaa - bbaaaabb.
\end{equation}
The motivation for this selection is that this channel is twice expanded X$_1$ channel and we may expect better performance than X$_1$ in lower frequency band.
\end{itemize}

The investigations for these selected TDI channels will be implemented by following four steps, 1) the TDI paths calculation using the algorithm in Section \ref{subsec:algorithm}, 2) the GW response analysis for TDI channels (in Section \ref{sec:response}), 3) the noise level evaluation of the channels (in Section \ref{sec:noise}), and 4) their sensitivities synthesis (in Section \ref{sec:sensitivity}).

\section{GW Response} \label{sec:response}

The response of a TDI channel to GW signal is the combination of the response in every single link. And the GW response formula for a single link has been specified in \cite{1975GReGr...6..439E,1987GReGr..19.1101W,Vallisneri:2007xa,Vallisneri:2012np}. We reiterate the response formulation as follow.

For a GW source located at a direction $(\lambda, \beta)$ in the SSB coordinates, where $\lambda $ and $\beta$ is the ecliptic longitude and latitude, the propagation vector $\hat{k} $ is
\begin{equation} \label{eq:source_vec}
 \hat{\bm{k}}  = -( \cos \lambda \cos \beta, \sin \lambda \cos \beta ,  \sin \beta ).
\end{equation} 
The plus and cross polarization tensors of the GW signal are
\begin{equation}
\begin{aligned}
{\bm{\mathrm{e}}}_{+} & \equiv \mathcal{O}_1 \cdot 
\begin{pmatrix}
1 & 0 & 0 \\
0 & -1 & 0 \\
0 & 0 & 0
\end{pmatrix}
\cdot \mathcal{O}^T_1 ,
\ \ 
{\bm{\mathrm{e}}}_{\times} & \equiv \mathcal{O}_1 \cdot 
\begin{pmatrix}
0 & 1 & 0\\
1 & 0 & 0 \\
0 & 0 & 0
\end{pmatrix}
\cdot \mathcal{O}^T_1,
\end{aligned}
\end{equation}
with
\begin{widetext}
\begin{equation}
\mathcal{O}_1 =
\begin{pmatrix}
\sin \lambda \cos \psi - \cos \lambda \sin \beta \sin \psi & -\sin \lambda \sin \psi - \cos \lambda \sin \beta \cos \psi & -\cos \lambda \cos \beta  \\
     -\cos \lambda \cos \psi - \sin \lambda \sin \beta \sin \psi & \cos \lambda \sin \psi - \sin \lambda \sin \beta \cos \psi & -\sin \lambda \cos \beta  \\
         \cos \beta \sin \psi & \cos \beta \cos \psi & -\sin \beta 
\end{pmatrix},
\end{equation}
where $\psi$ is the polarization angle. The GW response in the $l$th link in TDI paths from sender S/C$s$ to receiver S/C$r$ is
\begin{equation} \label{eq:resp_link}
\begin{aligned}
 y^{h}_{sr,l} (f, \Omega, \bm{r}_s, \bm{r}_r, l) =&  \frac{ (1 + \cos^2 \iota ) \hat{\bm{n}}_{sr} \cdot {\bm{\mathrm{e}}}_+ \cdot \hat{\bm{n}}_{sr} + i (- 2 \cos \iota ) \hat{\bm{n}}_{sr} \cdot {\bm{\mathrm{e}}}_\times \cdot \hat{\bm{n}}_{sr} }{4 (1 - \hat{\bm{n}}_{sr} \cdot \hat{\bm{k}} ) } 
\times \left[  e^{ 2 \pi i f ( \hat{\bm{k}} \cdot \bm{r}_s - \tau_s ) } -  e^{ 2 \pi i f ( \hat{\bm{k}} \cdot \bm{r}_r - \tau_r ) }  \right] ,
\end{aligned}
\end{equation}
\end{widetext}
where $\iota$ is the inclination angle of GW source, $\hat{\bm{n}}_{sr}$ is the unit vector from sender S/C$s$ to receiver S/C$r$, and $\bm{r}_{s/r}$ is the position of laser sender/receiver in the SSB coordinates as determined in the first step calculation. The $\hat{\bm{n}}_{sr}$ and $\bm{r}_{s/r}$ correspond to the values of $l$th link and the symbol $l$ is omitted in the right part of Eq. \eqref{eq:resp_link}.

As Eqs. \eqref{eq:source_vec}-\eqref{eq:resp_link} show, the response depends on four geometric angles $\Omega (\lambda, \beta, \psi, \iota)$, GW frequency $f$, time delay factors, and the positions of sender and receiver. Since the time delay and positions of S/C have been determined in the first step as the result in Table \ref{tab:TDI_X}, the response of TDI for a given $\Omega$ will be straightforward to calculate along with all TDI links,
\begin{equation} \label{eq:resp_TDI_FD}
 F^{h}_{ \rm TDI} (f, \Omega ) = \sum^{n}_{l = 1} \mathrm{sgn}( \tau_{l} - \tau_{l-1} ) y^{h}_{sr,l}(f, \Omega, \bm{r}_i, \bm{r}_j, l).
\end{equation}

\begin{figure}[htb]
\includegraphics[width=0.48\textwidth]{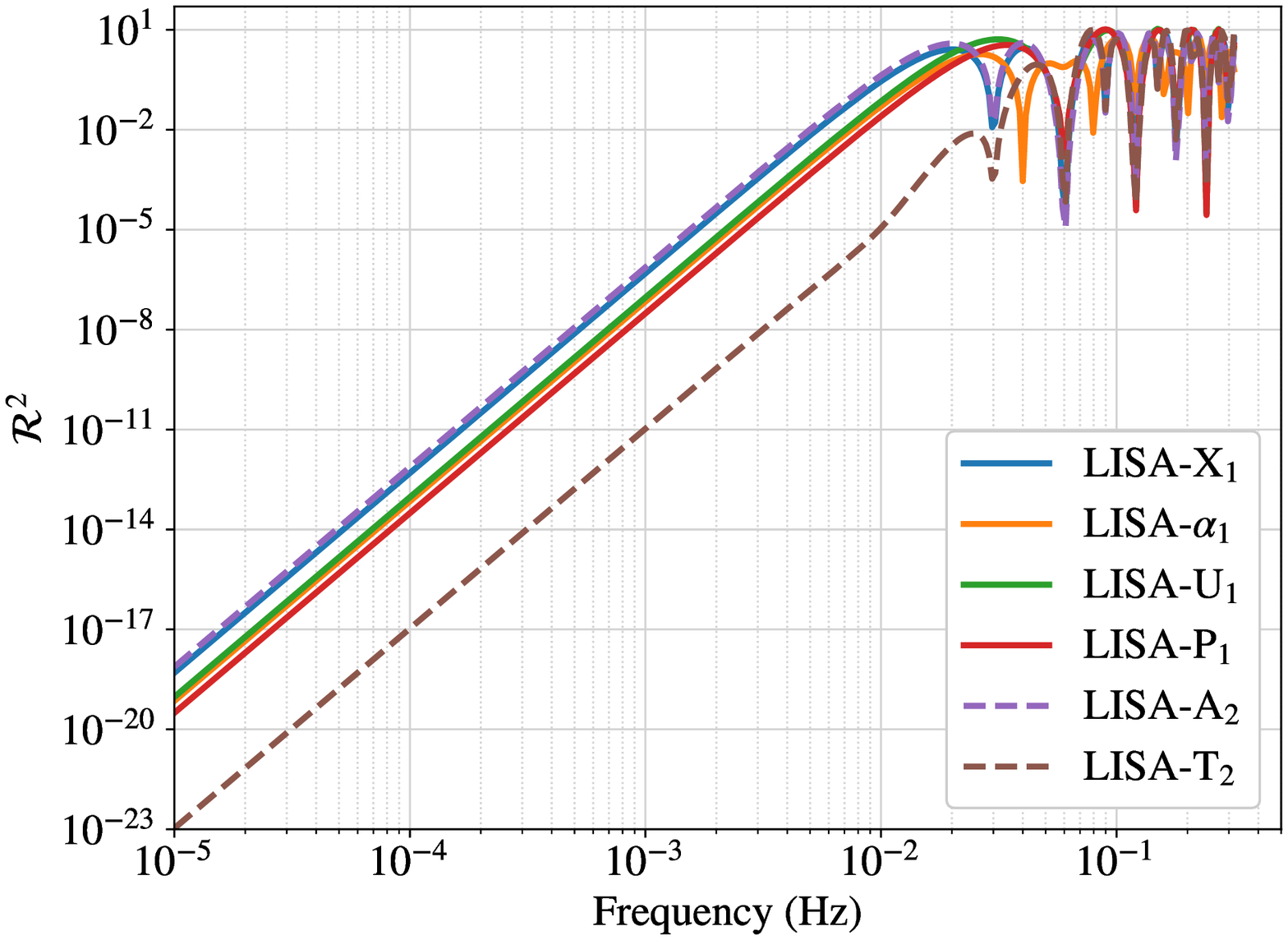}
\includegraphics[width=0.48\textwidth]{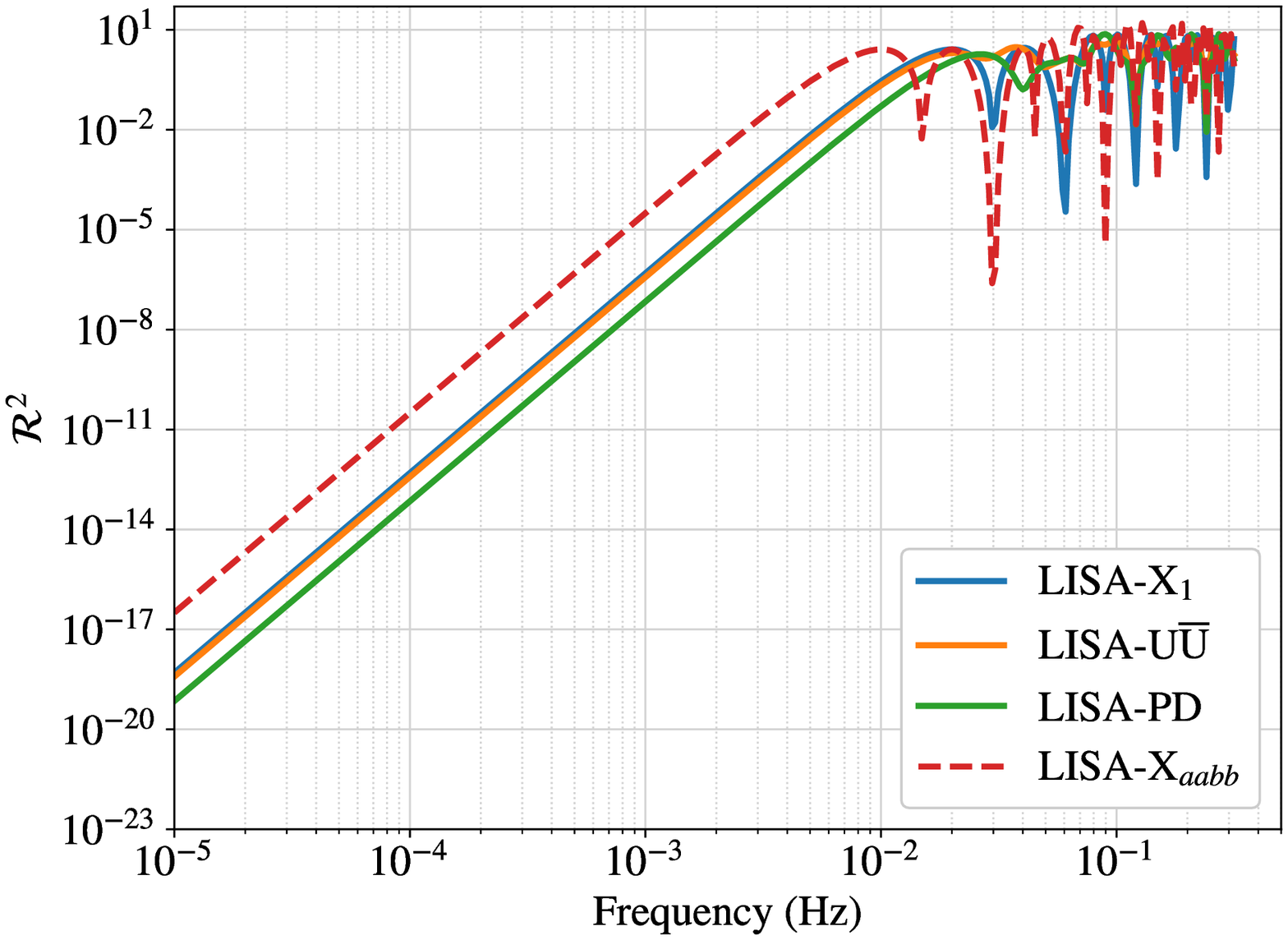}
\caption{\label{fig:TDI_resp_LISA} The median responses of TDI channels in the frequency spectrum over one year and $\Omega$(sky location $\lambda$ and $\beta$, polarization $\psi$ and inclination $\iota$).  The upper panel shows the first and second TDI groups, and the lower panel shows the third and fourth groups. The X$_1$ curve is treated as fiducial and shown also in the lower panel.}
\end{figure}

To evaluate the response for different $\Omega$ and frequency $f$ in a yearly orbit, we randomly sample $10^5$ sources in the $\Omega(\lambda, \beta, \psi, \iota)$ parameter space, and select the 26 time points in one year with a 14 days interval to calculate the responses in each TDI channel. This sampling method has been verified in \citet{Vallisneri:2012np} to achieve sufficient accuracy. And the average response of one TDI channel to a  monochromatic source in one year-observation will be
\begin{equation} \label{eq:resp_averaged_year}
 \mathcal{R}^2_{\rm TDI} (f, \Omega) = \frac{1}{T}  \int^{T}_{0} |F^{h}_{ \rm TDI} (f, \Omega)|^2  {\rm d} t,
\end{equation}
where $T$ is the observation time and is set to be one year. The median response to GW sources over sky location, polarization, and inclination is employed to represent the responsiveness of each channel, and the curves are shown in Fig. \ref{fig:TDI_resp_LISA}. The differences of GW response for TDI channels are the amplitudes at frequency region below 20 mHz and the spikes drop at their higher characteristic frequencies. The curve of the X$_1$ channel is treated as the fiducial response and shown in both upper and lower panels. In the lower frequency band, the response of X$_{aabb}$ is the highest in selected channels, and the response of T$_2$ is the lowest. 
The performance X$_{aabb}$ is supposed to be the result that its double round trip along each arm can cumulate more low frequency GW signal and its links are most in selected channels. 
For the response of T$_2$ channel, based on our investigation for T channel in \citet{Wang:2020a}, we infer that its low response is caused by cancellation from X$_1$, X$_2$ and X$_3$ equal part combination, and the inequality between arms would uplift its responsiveness.
For other channels, the A$_2$ channel has a slightly higher response than X$_1$, and the response from the rest channels are modestly lower than X$_1$. 

\section{Noises in TDI Channels} \label{sec:noise}

TDI is targeting to suppress laser frequency noise beneath the secondary noises. 
The first-generation TDI configurations can not sufficiently suppress the laser frequency noise for LISA as we investigated in \citet{Wang:2020fwa}. In this section, we investigate the residual laser noise and the secondary core noises (acceleration noise and optical path noise) levels in selected second-generation TDI channels. Other secondary noises (for instance, clock noise,  tilt-to-length noise) are supposed to be resolved by the new designs/methods \cite{Otto:2012dk,Otto:2015,Tinto:2018kij,Hartwig:2020tdu,Chwalla:2016bzk,Trobs:2017msu}.
By substituting Eqs. \eqref{eq:s_epsilon_tau_1} and \eqref{eq:s_epsilon_tau_2} into Eqs. \eqref{eq:eta} and \eqref{eq:TDI_algorithm} and summing up noises along the paths, we can obtain the noise level for a given TDI channel,
\begin{equation} \label{eq:TDI_algorithm}
 \mathrm{S_{n, TDI}} = \left| \sum^{n}_{l = 1} \mathrm{sgn}( \tau_{l} - \tau_{l-1} ) \tilde{\eta}_{ij,l}(t_m + \tau_{j,l} ) \right|^2,
\end{equation}
where $n$ is the number of links in a TDI channel, and $\tilde{\eta}$ is the amplitude spectral density of $\eta$. We decompose the laser frequency noise and secondary noise to show the impact of laser noise suppression and secondary noise level.

\subsection{Laser frequency noise} \label{subsec:laser_noise}

The laser noise terms after TDI combination could be obtained by substituting Eqs. \eqref{eq:s_epsilon_tau_1}-\eqref{eq:s_epsilon_tau_2} and \eqref{eq:eta} into \eqref{eq:TDI_combination}. For instance, the laser noise in the first-generation Michelson-X will be \cite{Wang:2020fwa},
\begin{equation}
\begin{aligned}
\mathrm{X}_\mathrm{laser} (t) = & C_{12} ( t_{m} + \tau_{1,0} ) - C_{12} (t_{m} + \tau_{1,8} ) 
\end{aligned}
\end{equation}
where $C_{12}$ is the noise from laser source on the optical bench S/C1 pointing to S/C2 (as described in Appendix \ref{sec:appendix_observables}), $\tau_{1,0}$ and $\tau_{1,8}$ are the first and last values of $\tau$ in Table \ref{tab:TDI_X}, respectively. And the mismatch between two laser paths is $\delta t = \tau_{1,0} - \tau_{1,8} $.
The amplitude of the Fourier components of residual laser noise would be \cite{Tinto:2020fcc},
\begin{equation} \label{eq:dC}
\begin{aligned}
 |  \widetilde{ \mathrm{X} }_\mathrm{laser} |  \simeq & 2 \pi f | \tau_{1,0} - \tau_{1,8}  | |\tilde{C}(f) | \\
  \simeq & 2 \pi f | \delta t | |\tilde{C}(f) |
 \end{aligned}
\end{equation}
where $\tilde{C}(f) \simeq 1 \times 10^{-13} / \sqrt{\rm Hz}$ is the one-sided square-root spectrum density of the Nd:YAG laser requirement for LISA \cite{2017arXiv170200786A}.

\begin{figure}[htb]
\includegraphics[width=0.48\textwidth]{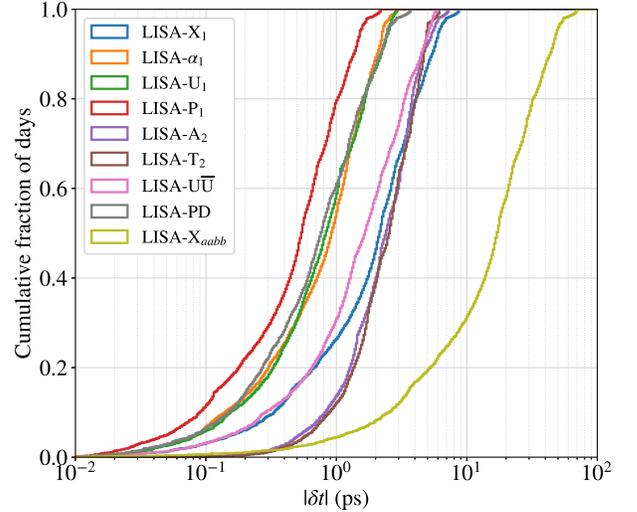}
\caption{\label{fig:TDI_dt_LISA} The cumulative histograms of path mismatches for selected TDI channels. By employing a set of 2200 days numerical orbit for LISA mission \cite{Wang:2017aqq,Wang:2020a}, the laser path mismatch of a TDI is calculated for each day (one point per day). The levels of path mismatch increases with longer time delay involved in TDI as reflected in Figs. \ref{fig:2nd_TDI_diagram} and \ref{fig:2nd_TDI_diagram_others} and Eq. \eqref{eq:X_aabb}.}
\end{figure}

By employing a set of 2200 days numerical orbit for LISA \cite{Wang:2017aqq,Wang:2020a}, the mismatching time for each TDI channel is calculated for each day by the algorithm in Section \ref{subsec:algorithm}, and their cumulative histograms are shown in Fig. \ref{fig:TDI_dt_LISA}.
By assuming the mismatch in 100 ns (30 m) is sufficient for laser noise cancellation for LISA \cite{2003PhRvD..67l2003T}, all the selected channels satisfy the requirement by several orders lower and can suppress laser noise effectively. The levels of path mismatch are varying with TDI channels, and we infer that the mismatch of a TDI channel increases with larger time span. And this inference could be reflected in the diagrams in Figs. \ref{fig:2nd_TDI_diagram} and \ref{fig:2nd_TDI_diagram_others} and Eq. \eqref{eq:X_aabb}: the X$_{aabb}$ channel with the largest mismatch has a longest time span ($16 L$), the mismatches of X$_1$ and U$\overline{\mathrm{U}}$ with $8L$ time range is larger than the channels expanded $6L$ ($\alpha_1$, U$_1$, and PD), and P$_1$ has the least mismatch for its $5L$ time span. 

\subsection{Secondary noise} \label{secsub:TDI_noise}

\begin{figure}[htb]
\includegraphics[width=0.49\textwidth]{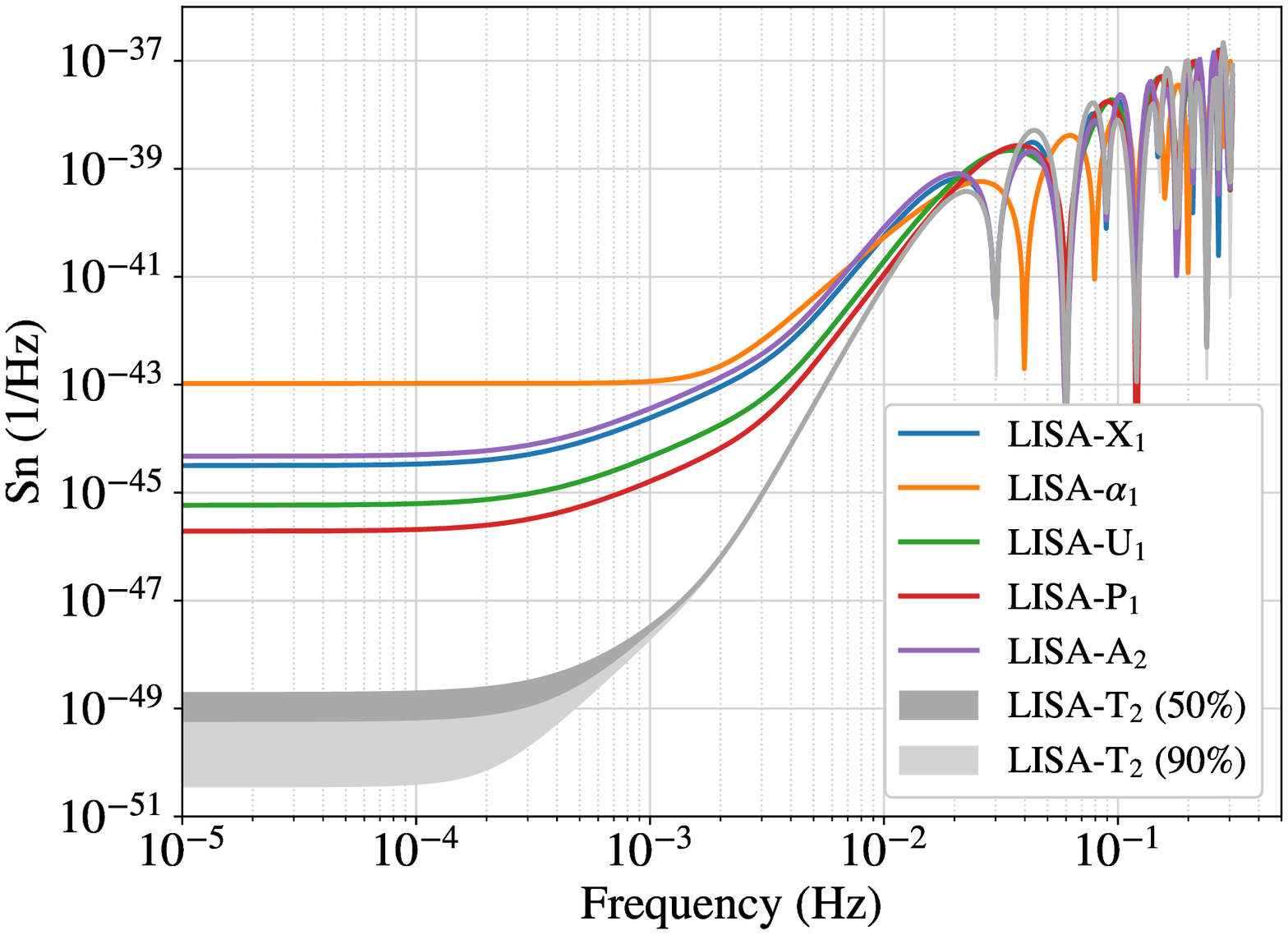}
\includegraphics[width=0.49\textwidth]{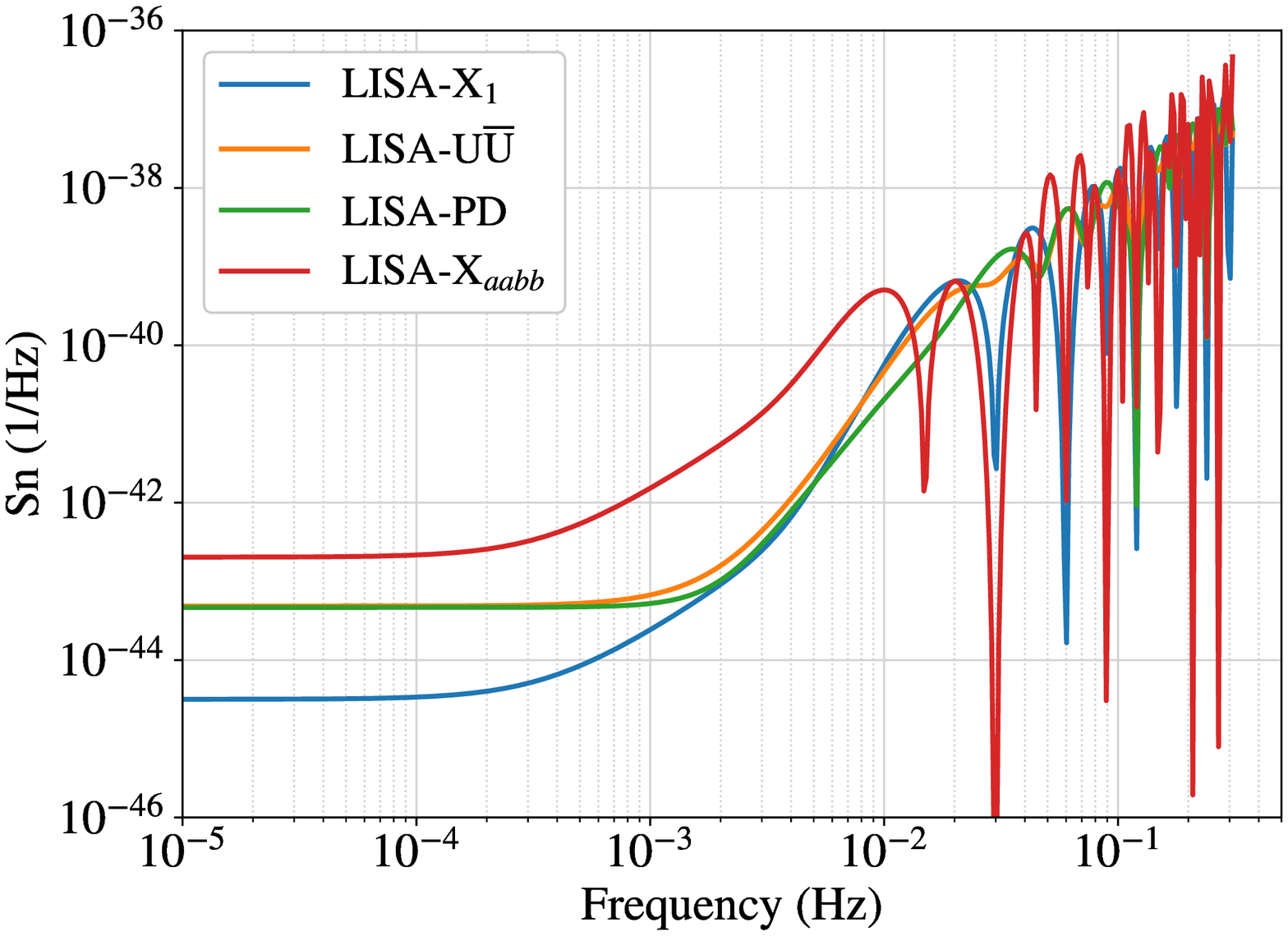}
\caption{\label{fig:Sn_secondary_noise_LISA} The secondary noise PSD of selected TDI channels in the frequency spectrum. The upper panel shows the noise PSD in the first group channels (X$_1$, $\alpha_1$, U$_1$, and P$_1$) and optimal channels (A$_2$ and T$_2$), and lower panel shows noise PSD in the channels U$\overline{\mathrm{U}}$, PD and X$_{aabb}$. The dark grey in upper panel shows the 50\% percentile highest noise level in T$_2$ channel in the first 300 mission days, and light grey together with dark grey show 90\% percentile.}
\end{figure}

After the laser frequency noise is substantially suppressed by the second-generation TDI, the secondary noises, especially acceleration noise and optical path noise become the dominant noise sources. By assuming that no correlation between the different test masses and optical benches, the noise level is evaluated by substituting the corresponding terms in Eqs. \eqref{eq:s_epsilon_tau_1} and \eqref{eq:s_epsilon_tau_2} into Eqs. \eqref{eq:eta} and \eqref{eq:TDI_algorithm}. Considering the upper limits of requirements for acceleration noise $\mathrm{ S_{n, acc} }$ and optical path noise $\mathrm{S_{n, op}}$ for LISA mission \cite{2017arXiv170200786A},
\begin{equation}
\begin{aligned}
& \sqrt{ \mathrm{ S_{n, acc} } } = 3 \times 10^{-15} \frac{\rm m/s^2}{\sqrt{\rm Hz}} \sqrt{1 + \left(\frac{0.4 {\rm mHz}}{f} \right)^2 }  \sqrt{1 + \left(\frac{f}{8 {\rm mHz}} \right)^4 }, \\
& \sqrt{ \mathrm{ S_{n, op} } } = 10 \times 10^{-12} \frac{\rm m}{\sqrt{\rm Hz}} \sqrt{1 + \left(\frac{2 {\rm mHz}}{f} \right)^4 }, \\
 \end{aligned}
\end{equation}
the noise levels for selected TDI channels are shown in Fig. \ref{fig:Sn_secondary_noise_LISA}. The upper panel shows the noise power spectrum density (PSD) of TDI channels in the first group (X$_1$, $\alpha_1$, U$_1$, and P$_1$) and optimal channels (A$_2$ and T$_2$), and the lower panel shows noise PSD in TDI channels U$\overline{\mathrm{U}}$, PD, and X$_{aabb}$. By employing an unequal-arm numerical calculation, we find that the noise PSD of T$_2$ channel has a visible variance with time (or with the inequality of arm lengths) for frequencies lower than 1 mHz, and this phenomenon also appeared in the PSD calculation for the first-generation T channel as shown in \citet{Wang:2020fwa} . The dark grey in the upper panel shows the 50\% percentile highest noise level in the first 300 days, and light grey together with dark grey show 90\% percentile noise level in 300 days for T$_2$ channel.

For the TDI channels in the first group, their PSD is approximately proportional to $ 4 \sin^2(n \pi f L)$ of their corresponding first-generation TDI PSDs, where $n$ is the number of time shifted arm length $L$ between two first-generation TDI channels as shown in Eq. \eqref{eq:regular_TDI} and Fig. \ref{fig:2nd_TDI_diagram}. Therefore, the first dropping spike appears at $f = 1/(4L) \simeq 0.03$ Hz for X$_1$, A$_2$ and T$_2$, $f = 1/(3L) \simeq 0.04$ Hz for $\alpha_1$, $f = 1/(2L) \simeq 0.06$ Hz for U$_1$ and P$_1$, and $f = 1/(8L) \simeq 0.015$ Hz for X$_{aabb}$ channel. The rule is not adaptable for the U$\overline{\mathrm{U}}$ and PD channels which utilize two different first-generation channels.

\section{Sensitivities of TDI channels} \label{sec:sensitivity}

Based on the response and noise level for a GW source with $\Omega$ (ecliptic longitude $\lambda$, latitude $\beta$, polarization $\psi$, inclination $\iota$), the optimal signal-to-noise ratio (SNR), $\rho_\mathrm{opt}$, for one mission will be the joint SNRs from three optimal TDI channels (for instance, $\mathrm{A_2, E_2, T_2}$) \cite{Tinto:2020fcc},
\begin{equation} \label{eq:SNR}
\begin{aligned}
\rho^2_\mathrm{opt} = &  \int^{f_\mathrm{max}}_{f_\mathrm{min}} \sum_{\mathrm{A}_2, \mathrm{E}_2, \mathrm{T}_2} \frac{ | F^h_\mathrm{TDI} ( f, L, \Omega ) \ast \tilde{h}(f) |^2 }{ {\mathrm{S}_\mathrm{n,TDI}(f, L )} } \mathrm{d} f,
\end{aligned}
\end{equation}
where $\tilde{h}(f)$ is the GW signal in frequency domain.
The instantaneous optimal sensitivity, $\mathrm{S}_\mathrm{opt} (f, L, \Omega)$, could be derived from Eq. \eqref{eq:SNR} as \cite{Tinto:2020fcc}
\begin{equation}
\begin{aligned}
\mathrm{S}_{\mathrm{opt}}(f, L, \Omega) &= \left[ \sum_{\mathrm{A_2, E_2, T_2}} \frac{  |F^h_{\mathrm{TDI} }( f, L, \Omega) |^2  }{\mathrm{S}_{\mathrm{n,TDI}} ( f, L )} \right]^{-1},
\end{aligned}
\end{equation}
and $\mathrm{S}_{\mathrm{opt}}$ represents the strain precision which could be measured when the six laser links of mission are functional. Considering the constellation orbital motions and time-varying arm lengths, $F^h_{\mathrm{TDI} }( f, L, \Omega) $ and $S_\mathrm{n,TDI} (f, L)$ will change with mission time, especial for the $\mathrm{T_2}$ channel. Therefore, the yearly averaged sensitivity for a source with $\Omega$ could evaluated as,
\begin{equation}
\begin{aligned}
\overline{\mathrm{S}}_{\mathrm{opt}}(f, \Omega) &= \left[ \sum_{\mathrm{A_2, E_2, T_2}} \left\langle \frac{  | F^h_{\mathrm{TDI} }( f, L, \Omega) |^2  }{\mathrm{S}_{\mathrm{n,TDI}} ( f, L )} \right\rangle_\mathrm{1yr} \right]^{-1} \\
& = \left[  \sum_{\mathrm{A_2, E_2, T_2}} \frac{1}{T} \int^T_0 \frac{ |F^h_{\mathrm{TDI} }(f, L, \Omega) |^2 }{{\mathrm{S}_{\mathrm{n, TDI} }( f, L )} } \mathrm{d} t  \right]^{-1}.
\end{aligned}
\end{equation}
And the yearly averaged sensitivity for a single TDI channel will be
\begin{equation}
\overline{\mathrm{S}}_{\mathrm{TDI}} (f, \Omega) =  \left[  \frac{1}{T} \int^T_0 \frac{ | F^h_{\mathrm{TDI}}(f, L, \Omega ) |^2}{\mathrm{S}_{\mathrm{n, TDI}}(f, L )} \mathrm{d} t \right]^{-1}.
\end{equation}  
To make following content more readable, we clarify that $\mathrm{S}_{\mathrm{n, TDI}}$ denotes the noise PSD from the instruments/measurements in a TDI channel, and $\mathrm{S}_{\mathrm{TDI}}$ is the sensitivity of a TDI channel which is the GW response weighted $\mathrm{S}_{\mathrm{n, TDI}}$. 

\begin{figure*}[htb]
\includegraphics[width=0.49\textwidth]{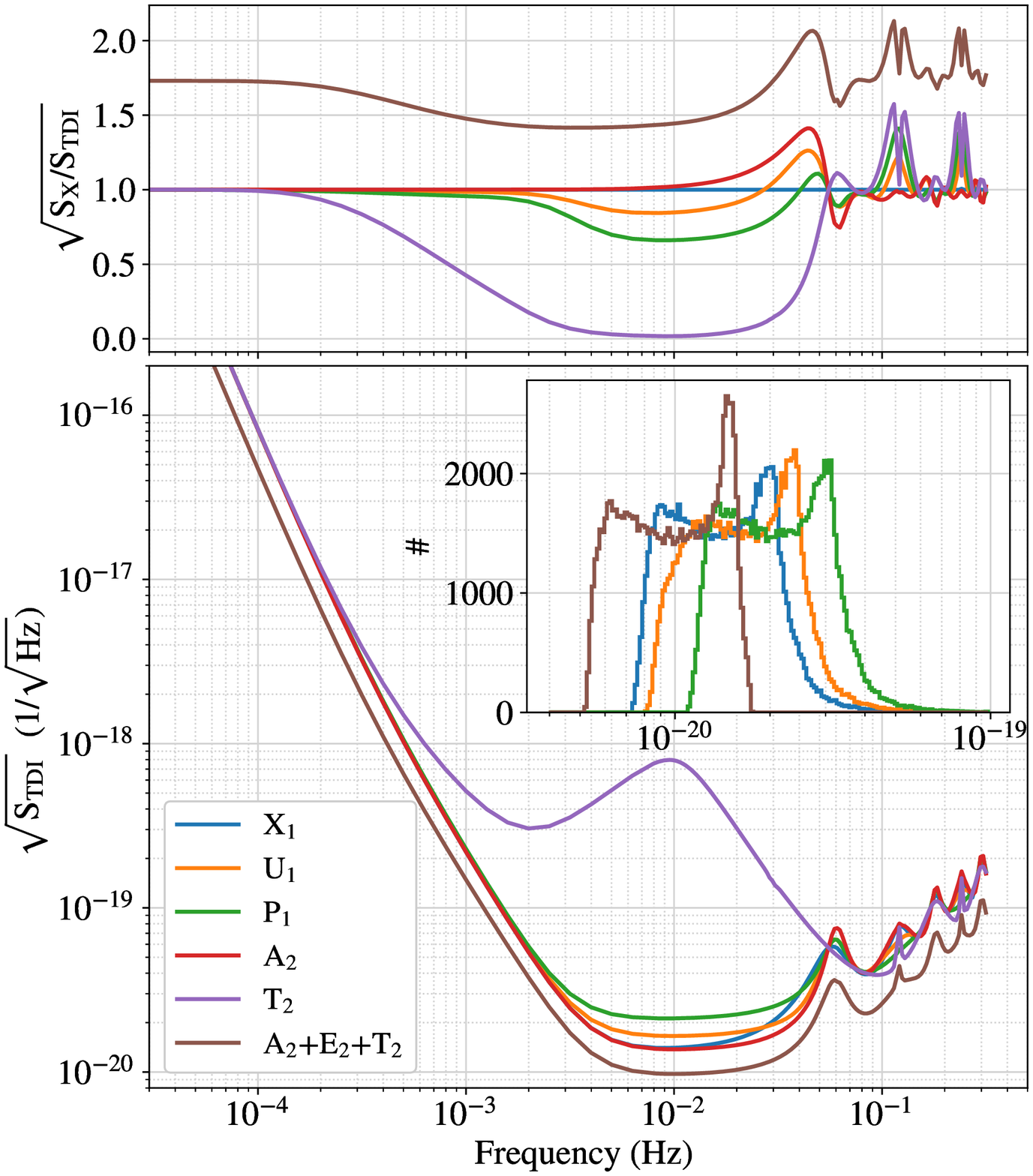}
\includegraphics[width=0.49\textwidth]{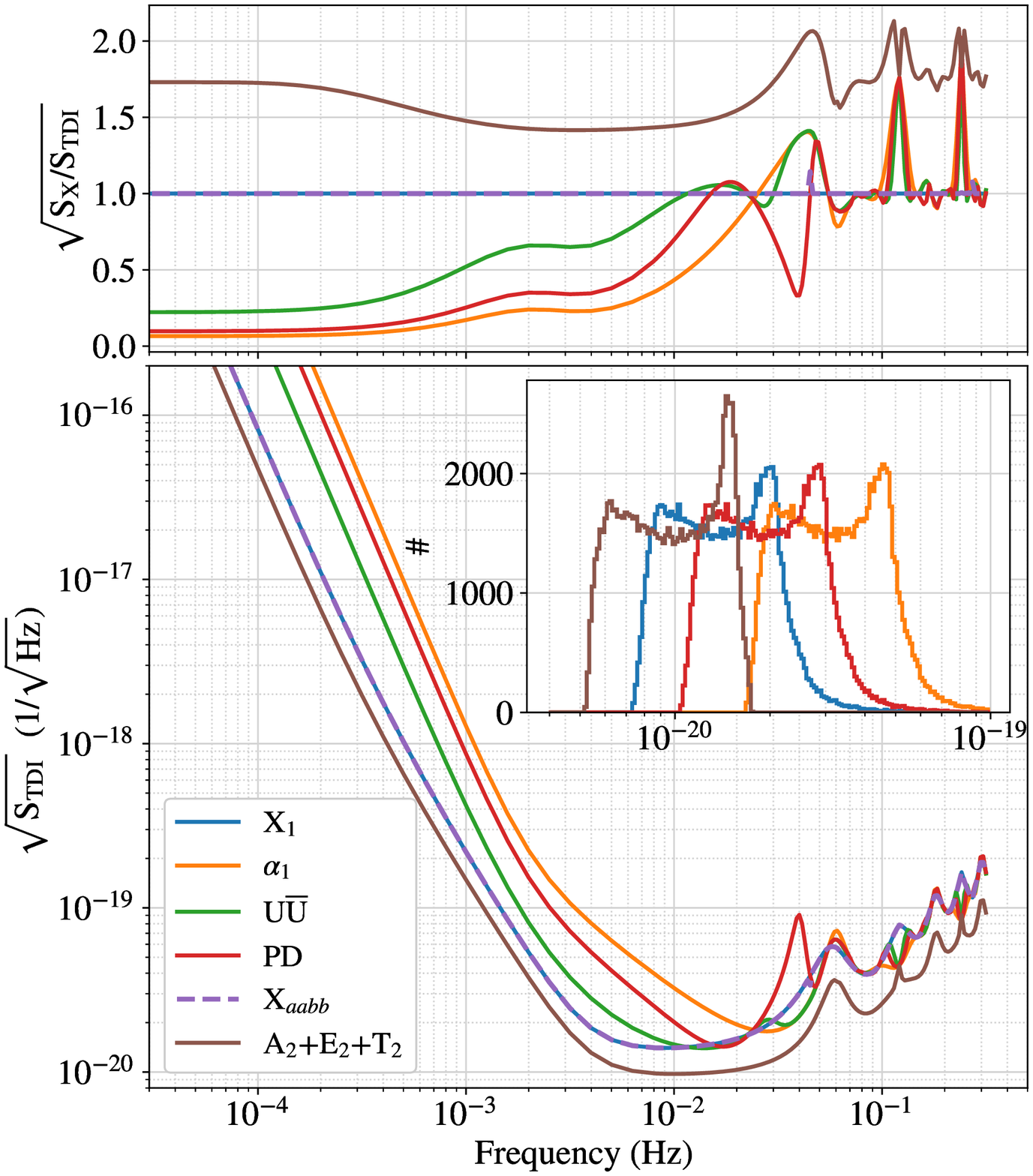}
\caption{\label{fig:sensitivity_curves} The yearly median sensitivity over $\Omega(\lambda, \beta, \psi, \iota)$ for selected TDI channels. 
The left panel shows the curves for channels X$_1$, U$_1$, and P$_1$ and optimal channels (A$_2$ and T$_2$), and the right panel shows channels $\alpha_1$, U$\overline{\mathrm{U}}$, PD and X$_{aabb}$ as well as X$_{1}$. The joint sensitivity $\mathrm{A_2+E_2+T_2}$ is plotted in both panels for comparison.
The upper panels show the sensitivity ratios between fiducial first-generation laser noise free Michelson-X and selected channels, $\sqrt{ \mathrm{S}_{\mathrm{X}} / \mathrm{S}_{\mathrm{TDI}} }$. The inset plots in lower panels show the histograms of sensitivities at 10 mHz for multiple channels. (Note that, these sensitivities are averaged over the sky location, polarization and inclination $\Omega$($\lambda, \beta, \psi, \iota$), the sensitivity over sky location and polarization ($\lambda, \beta, \psi, \iota=0$) will be lower than these results by a factor of $\frac{8}{5\pi} \simeq 0.51$)
}
\end{figure*}

\begin{figure}[htb]
\includegraphics[width=0.49\textwidth]{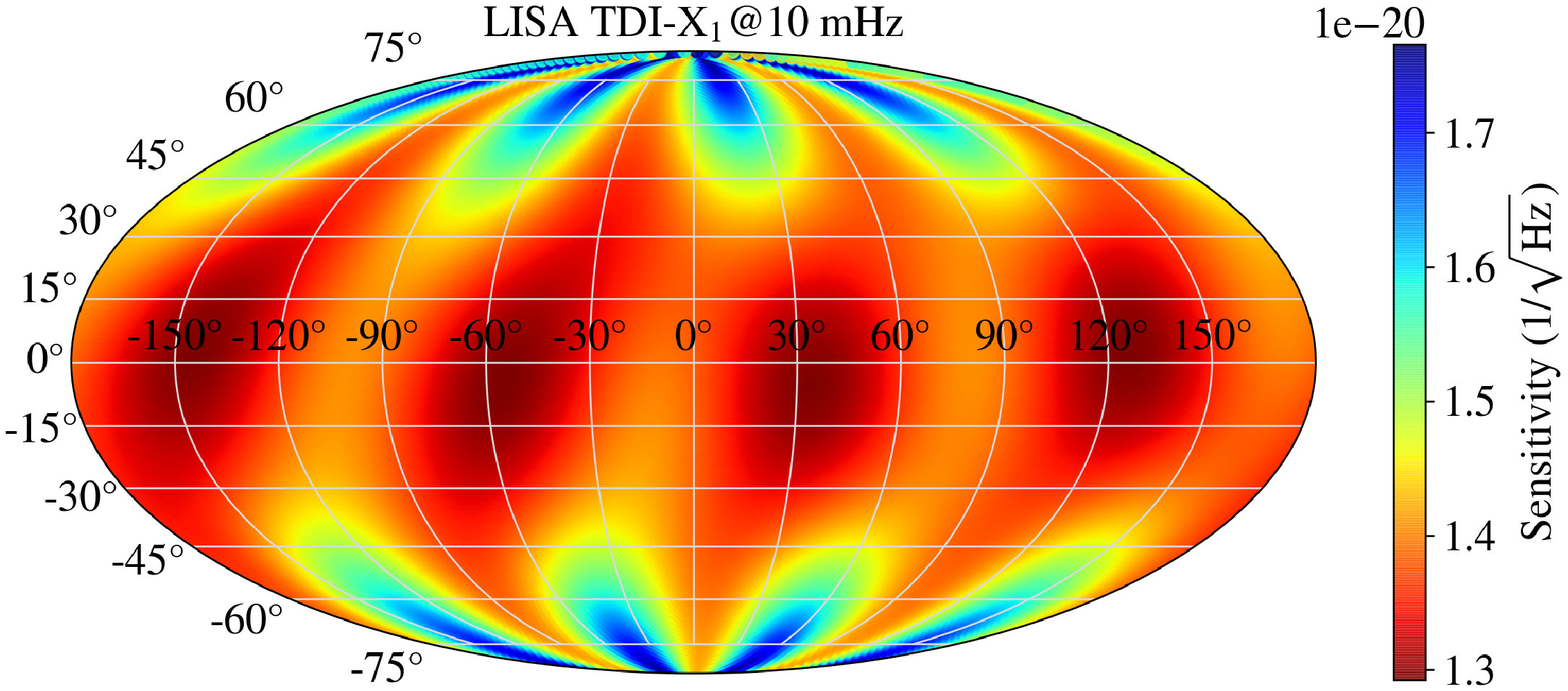}
\includegraphics[width=0.49\textwidth]{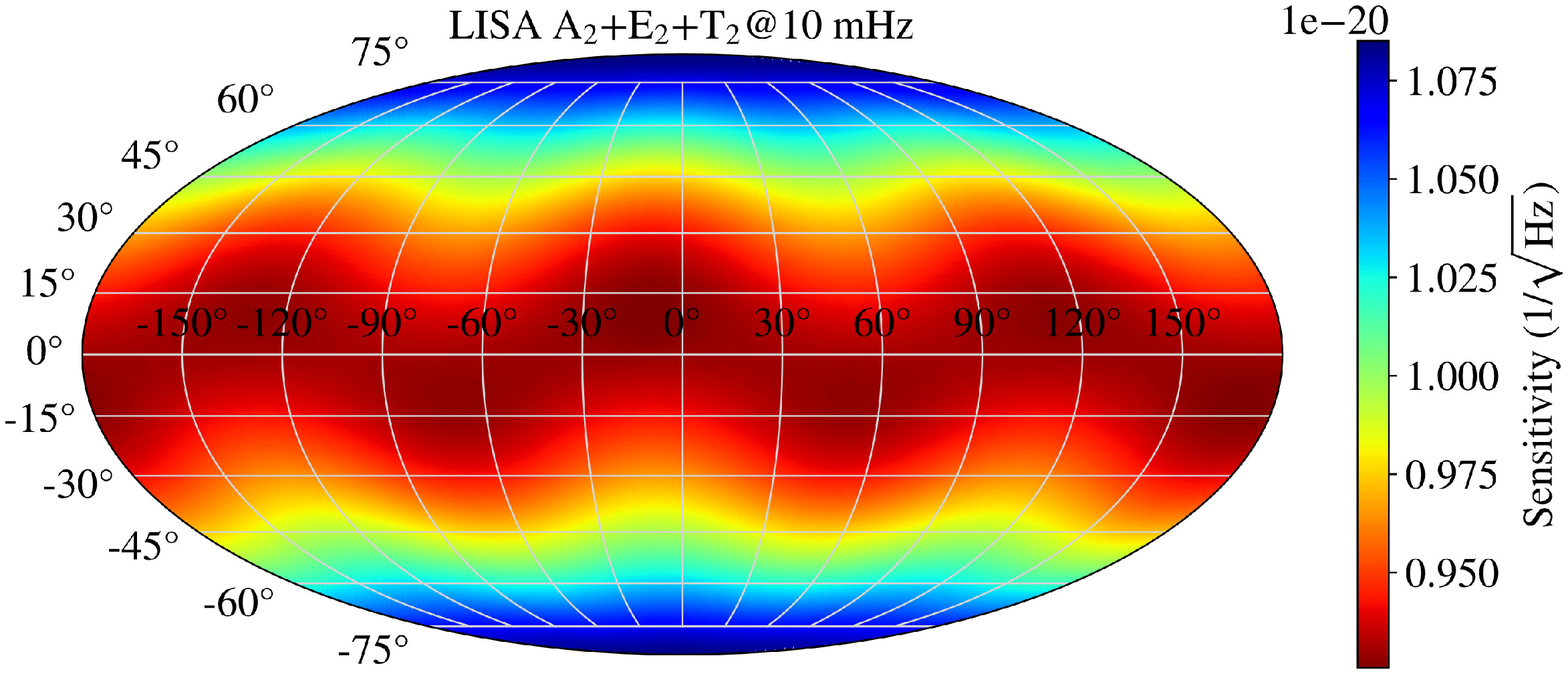}
\caption{\label{fig:sensitivity_mollweide} The yearly averaged sensitivity distribution on sky map for X$_1$(upper panel) and the joint optimal A$_2$, E$_2$ and T$_2$ (lower panel) channels at 10 mHz in the solar system barycentric ecliptic coordinates. The sensitivity is calculated by fixing the polarization ($\psi = \pi/6$) and inclination ($\iota = \pi/3$). }
\end{figure}

As we can expect, the sensitivity of a TDI channel will be different for various $\Omega(\lambda, \beta, \psi, \iota)$, the histograms of sensitivities at 10 mHz for multiple channels are shown in inset plots of lower panels of Fig. \ref{fig:sensitivity_curves}. Comparing to the single channel with a longer tail at the sensitivity becoming worse, the histogram of joint $\mathrm{A_2+E_2+T_2}$ channel narrows down the sensitivity range which indicates the more homogeneous space coverage. The most sensitive direction is around the ecliptic plane, while the insensitive direction is around the ecliptic polar directions in one-year observation as shown in Fig. \ref{fig:sensitivity_mollweide}.
The further calculation shows that the mean value of sensitivity over $\Omega$ is $\sim$1.14 times worse than its median value except $\mathrm{A_2+E_2+T_2}$'s $\sim$1.1. 
The curve of median values is employed to represent the performance of each TDI channel over $\Omega$ parameter space, and the curves of selected channels are shown in Fig. \ref{fig:sensitivity_curves}.
Note that, this sensitivity is averaged over sky location, polarization and inclination ($\lambda, \beta, \psi, \iota$), the sky location and polarization ($\lambda, \beta, \psi,\iota=0$) averaged sensitivity will be lower than these results by a factor of $\frac{8}{5\pi} \simeq 0.51$.

The curves of yearly averaged median sensitivity over $\Omega$ for channels (X$_1$, $\alpha_1$, U$_1$, P$_1$, A$_2$, and T$_2$) are shown in the left panel of Fig. \ref{fig:sensitivity_curves}, and curves for other channels (X$_{1}$, $\alpha_1$, U$\overline{\mathrm{U}}$, PD, and X$_{aabb}$) are shown in the right panel, as well as the curves of X$_1$ and $\mathrm{A_2+E_2+T_2}$ in both panels for comparison. As we can see from the left plot, compared to the first-generation TDI channels shown in \cite{Wang:2020a,Wang:2020fwa}, the sensitivities of channels (X$_1$, U$_1$ and P$_1$, A$_2$ and T$_2$) are expected to be equal to their corresponding first-generation TDI channels when the laser frequency noise is not considered. The sensitivity of the T$_2$ channel is irregular and should be the result of unequal arm configuration as we investigated in \cite{Wang:2020a,Wang:2020fwa}. The joint $\mathrm{A_2+E_2+T_2}$ channel not only improves the sensitivity of X$_1$ by a factor of $\sqrt{2}$ to $\sqrt{3}$ for frequencies lower than 30 mHz, and by a factor of 2 at some higher frequencies as shown in the upper panel; it also achieves a better sky coverage as shown by the histograms in the inset plots.

The major differences of sensitivities in the right panel of Fig. \ref{fig:sensitivity_curves} are in frequency range lower than 20 mHz. The sensitivity of $\alpha_1$ channel is the worst in selected channels at lower frequencies. However, it can reach a relatively good level at some higher frequencies (e.g. 30 mHz, 120 mHz). The PD channel has the second worse sensitivity at low frequencies, and reach $\sim$2 times better than X$_1$ at frequencies 0.12 Hz and 0.24 Hz. The U$\overline{\mathrm{U}}$ channel has a worse sensitivity than X$_1$ channel in lower frequency band, and has better sensitivity at some frequencies. The sensitivity of X$_{aabb}$ is identical to X$_1$ even its response in lower frequency band is higher than X$_1$, the higher noise level counteracts the advantage in response. And we infer that all Michelson-like TDI combinations have identical sensitivity since the GW response and noise level are proportional.

On the other hand, to illustrate the sensitivity variation with the sky locations of GW sources, with the fixed polarization and inclination ($\psi = \pi/6$, $\iota = \pi/3$), the yearly averaged sensitivity of X$_1$ and joint $\mathrm{A_2+E_2+T_2}$ channels at 10 mHz are shown in Fig. \ref{fig:sensitivity_mollweide}. As expected, with the orbital motion of 60$^\circ$ tilled array, the sources located around the ecliptic plane could be observed with better antenna pattern modulation and hence better sensitivity, and polar directions are observed with worst sensitivities. Another factor is that the sources at the same latitude could not be observed with equal sensitivity as shown in the upper panel for X$_1$ channel, and this should be due to the geometric angles between polarization and selected interferometric arms and their yearly orbital motion.
And without considering the variations of polarization and inclination, the sensitivity range for the insensitive polar directions and sensitive ecliptic plane directions becomes more concentrated comparing to the results considering the $\Omega$ four parameters.

\section{Conclusions} \label{sec:conclusions}

In this work, we introduce a generic algorithm to investigate the performance of TDI. By employing a numerical mission orbit for LISA, as the first step, the algorithm determines the time delays and S/C positions in a realistic dynamical case. And then the algorithm can numerically evaluate the GW response, PSD of noise sources, and sensitivity for a TDI channel based on the obtained time delays and S/C positions. As assistance to sequence the links in TDI, a S/C layout-time delay diagram is developed to streamline the calculation procedures in the algorithm. And the algorithm should be feasible for any TDI observable and other missions employing the TDI technology.

We select 11 second-generation TDI channels constructed from four approaches and implement our algorithm for their performance investigations. Based on the numerical results, the interference paths of selected TDI channels are well matched and the laser frequency noise should be sufficiently suppressed beneath the secondary noise.
Without considering laser frequency noise and only including secondary core noises (acceleration noise and optical path noise), the second-generation TDI channels composed from two same first-generation channels would have equivalent sensitivities as their corresponding first-generation channels. This is also applicable to the optimal TDI channels (A$_2$, E$_2$, and T$_2$) combined from the second-generation Michelson channels (X$_1$, X$_2$, and X$_3$) compared to the first-generation optimal channels (A, E and T). The joint $\mathrm{A_2+E_2+T_2}$ channel would have $\sqrt{2}$ to 2 times better sensitivity than X$_1$ channel in full frequency band, and the joint observation could also cover all sky directions with more even sensitivity. In one-year's observation, the sensitive direction is around the ecliptic plane, while the insensitive directions are around the ecliptic polar.

The Sagnac $\alpha_1$ together with PD and U$\overline{\mathrm{U}}$ channels have the worse sensitivity at low-frequency band compared to the fiducial X$_1$ channel, although they could have better sensitivities at multiple characteristic higher frequencies.
For the Michelson-like channels which employing four links from two arms, X, X$_1$, and X$_{aabb}$ are investigated, and their sensitivities are identical when the secondary core noises are considered. Even X$_{aabb}$ channel improves the GW response in the low-frequency band, the noise level uplifted due to the path combination counteracts the increase of the response. And all Michelson-like TDI channels are inferred to have equal sensitivity.

\begin{acknowledgments}
This work was supported by NSFC No. 12003059 and No. 11773059, Key Research Program of Frontier Sciences, Chinese Academy of Science, No. QYZDB-SSW-SYS016 and the Strategic Priority Research Program of the Chinese Academy of Sciences under grant Nos. XDA1502070102, XDA15020700, XDB21010100, XDB23030100 and XDB23040000. and by the National Key Research and Development Program of China under Grant Nos. 2016YFA0302002 and 2017YFC0601602. This work made use of the High Performance Computing Resource in the Core Facility for Advanced Research Computing at Shanghai Astronomical Observatory.
\end{acknowledgments}

\appendix
\section{Appendix}
\subsection{The observables on optical benches} \label{sec:appendix_observables}

Observables $s_{ji}$, $\varepsilon_{ij}$ and $\tau_{ij}$ for $j=$S/C2 $\rightarrow $ $i=$ S/C1, S/C3$\rightarrow$S/C2 and S/C1$\rightarrow $S/C3 )
\begin{equation} \label{eq:s_epsilon_tau_1}
\begin{aligned}
   s_{ji} = & y^h_{ji}:h + \mathcal{D}_{ji} C_{ji}(t) - C_{ij}(t) \\ &+ \mathcal{D}_{ji} N^{\rm OB}_{ji}(t) - N^{\rm OB}_{ij}(t) + n^{\rm op}_{ij}(t), \\
   \varepsilon_{ij} = & C_{ik}(t) - C_{ij}(t) + 2 n^{\rm acc}_{ij}(t) - 2 N^{\rm OB}_{ij}(t), \\
   \tau_{ij} = & C_{ik}(t) - C_{ij}(t) , 
\end{aligned}
\end{equation}
and observables $s_{ij}$, $\varepsilon_{ij}$ and $\tau_{ij}$ (for $1 \rightarrow 2$, $2 \rightarrow 3$ and $3 \rightarrow 1$ )
\begin{equation} \label{eq:s_epsilon_tau_2}
\begin{aligned}
   s_{ji} &=  y^h_{ji}:h + \mathcal{D}_{ji} C_{ji}(t) - C_{ij}(t) \\ & - \mathcal{D}_{ji} N^{\rm OB}_{ji}(t) + N^{\rm OB}_{ij}(t) + n^{\rm op}_{ij}(t), \\
   \varepsilon_{ij} &= C_{ik}(t) - C_{ij}(t) - 2 n^{\rm acc}_{ij}(t) + 2 N^{\rm OB}_{ij}(t), \\
   \tau_{ij} &= C_{ik}(t) - C_{ij}(t).
\end{aligned}
\end{equation}
The symbols are specified as follows.
\begin{itemize}
\item $y^h_{ji}$ is the response function to the GW signal $h$.
\item $C_{ij}$ denotes laser noise on the optical bench in S/C$i$ pointing to S/C$j$.
\item $N^{\mathrm{OB}}_{ij}$ is the effect from displacement along the arm $L_{ji}$ for the optical bench on S/C$i$ pointing to S/C$j$.
\item $L_{ij}$ is the arm length or propagation time from S/C$i$ to $j$ which includes the relativistic delay caused by gravitational field. The $L_{ij}$ and $ L_{ji}$ are treated as unequal in this dynamical scenario, and the calculation is described by Eq. \eqref{eq:L_ij}.
\item $n^{\mathrm{op}}_{ij}$ represents the optical path noise on the S/C$i$ pointing to $j$.
\item $n^{\mathrm{acc}}_{ij}$ denotes the acceleration noise from test mass on the S/C$i$ pointing to $j$.
\end{itemize}

\nocite{*}
\bibliography{apsref}

\providecommand{\noopsort}[1]{}\providecommand{\singleletter}[1]{#1}%
\begin{thebibliography}{57}%
\makeatletter
\providecommand \@ifxundefined [1]{%
 \@ifx{#1\undefined}
}%
\providecommand \@ifnum [1]{%
 \ifnum #1\expandafter \@firstoftwo
 \else \expandafter \@secondoftwo
 \fi
}%
\providecommand \@ifx [1]{%
 \ifx #1\expandafter \@firstoftwo
 \else \expandafter \@secondoftwo
 \fi
}%
\providecommand \natexlab [1]{#1}%
\providecommand \enquote  [1]{``#1''}%
\providecommand \bibnamefont  [1]{#1}%
\providecommand \bibfnamefont [1]{#1}%
\providecommand \citenamefont [1]{#1}%
\providecommand \href@noop [0]{\@secondoftwo}%
\providecommand \href [0]{\begingroup \@sanitize@url \@href}%
\providecommand \@href[1]{\@@startlink{#1}\@@href}%
\providecommand \@@href[1]{\endgroup#1\@@endlink}%
\providecommand \@sanitize@url [0]{\catcode `\\12\catcode `\$12\catcode
  `\&12\catcode `\#12\catcode `\^12\catcode `\_12\catcode `\%12\relax}%
\providecommand \@@startlink[1]{}%
\providecommand \@@endlink[0]{}%
\providecommand \url  [0]{\begingroup\@sanitize@url \@url }%
\providecommand \@url [1]{\endgroup\@href {#1}{\urlprefix }}%
\providecommand \urlprefix  [0]{URL }%
\providecommand \Eprint [0]{\href }%
\providecommand \doibase [0]{https://doi.org/}%
\providecommand \selectlanguage [0]{\@gobble}%
\providecommand \bibinfo  [0]{\@secondoftwo}%
\providecommand \bibfield  [0]{\@secondoftwo}%
\providecommand \translation [1]{[#1]}%
\providecommand \BibitemOpen [0]{}%
\providecommand \bibitemStop [0]{}%
\providecommand \bibitemNoStop [0]{.\EOS\space}%
\providecommand \EOS [0]{\spacefactor3000\relax}%
\providecommand \BibitemShut  [1]{\csname bibitem#1\endcsname}%
\let\auto@bib@innerbib\@empty
\bibitem [{\citenamefont {Abbott}\ \emph
  {et~al.}(2016{\natexlab{a}})\citenamefont {Abbott} \emph
  {et~al.}}]{Abbott:2016blz}%
  \BibitemOpen
  \bibfield  {author} {\bibinfo {author} {\bibfnamefont {B.~P.}\ \bibnamefont
  {Abbott}} \emph {et~al.} (\bibinfo {collaboration} {{LIGO Scientific
  Collaboration and Virgo Collaboration}}),\ }\bibfield  {title} {\bibinfo
  {title} {{Observation of Gravitational Waves from a Binary Black Hole
  Merger}},\ }\href {https://doi.org/10.1103/PhysRevLett.116.061102} {\bibfield
   {journal} {\bibinfo  {journal} {Phys. Rev. Lett.}\ }\textbf {\bibinfo
  {volume} {116}},\ \bibinfo {pages} {061102} (\bibinfo {year}
  {2016}{\natexlab{a}})},\ \bibinfo {note} {and references therein},\ \Eprint
  {https://arxiv.org/abs/1602.03837} {arXiv:1602.03837 [gr-qc]} \BibitemShut
  {NoStop}%
\bibitem [{\citenamefont {Abbott}\ \emph
  {et~al.}(2016{\natexlab{b}})\citenamefont {Abbott} \emph
  {et~al.}}]{TheLIGOScientific:2016pea}%
  \BibitemOpen
  \bibfield  {author} {\bibinfo {author} {\bibfnamefont {B.~P.}\ \bibnamefont
  {Abbott}} \emph {et~al.} (\bibinfo {collaboration} {{LIGO Scientific
  Collaboration and Virgo Collaboration}}),\ }\bibfield  {title} {\bibinfo
  {title} {{Binary Black Hole Mergers in the first Advanced LIGO Observing
  Run}},\ }\href {https://doi.org/10.1103/PhysRevX.6.041015,
  10.1103/PhysRevX.8.039903} {\bibfield  {journal} {\bibinfo  {journal} {Phys.
  Rev. X}\ }\textbf {\bibinfo {volume} {6}},\ \bibinfo {pages} {041015}
  (\bibinfo {year} {2016}{\natexlab{b}})},\ \bibinfo {note} {[erratum: Phys.
  Rev.X 8,no.3,039903(2018)]},\ \Eprint {https://arxiv.org/abs/1606.04856}
  {arXiv:1606.04856 [gr-qc]} \BibitemShut {NoStop}%
\bibitem [{\citenamefont {Abbott}\ \emph {et~al.}(2017)\citenamefont {Abbott}
  \emph {et~al.}}]{TheLIGOScientific:2017qsa}%
  \BibitemOpen
  \bibfield  {author} {\bibinfo {author} {\bibfnamefont {B.~P.}\ \bibnamefont
  {Abbott}} \emph {et~al.} (\bibinfo {collaboration} {{LIGO Scientific
  Collaboration and Virgo Collaboration}}),\ }\bibfield  {title} {\bibinfo
  {title} {{GW170817: Observation of Gravitational Waves from a Binary Neutron
  Star Inspiral}},\ }\href {https://doi.org/10.1103/PhysRevLett.119.161101}
  {\bibfield  {journal} {\bibinfo  {journal} {Phys. Rev. Lett.}\ }\textbf
  {\bibinfo {volume} {119}},\ \bibinfo {pages} {161101} (\bibinfo {year}
  {2017})},\ \Eprint {https://arxiv.org/abs/1710.05832} {arXiv:1710.05832
  [gr-qc]} \BibitemShut {NoStop}%
\bibitem [{\citenamefont {Abbott}\ \emph {et~al.}(2019)\citenamefont {Abbott}
  \emph {et~al.}}]{LIGOScientific:2018mvr}%
  \BibitemOpen
  \bibfield  {author} {\bibinfo {author} {\bibfnamefont {B.~P.}\ \bibnamefont
  {Abbott}} \emph {et~al.} (\bibinfo {collaboration} {{LIGO Scientific
  Collaboration and Virgo Collaboration}}),\ }\bibfield  {title} {\bibinfo
  {title} {{GWTC-1: A Gravitational-Wave Transient Catalog of Compact Binary
  Mergers Observed by LIGO and Virgo during the First and Second Observing
  Runs}},\ }\href {https://doi.org/10.1103/PhysRevX.9.031040} {\bibfield
  {journal} {\bibinfo  {journal} {Phys. Rev. X}\ }\textbf {\bibinfo {volume}
  {9}},\ \bibinfo {pages} {031040} (\bibinfo {year} {2019})},\ \Eprint
  {https://arxiv.org/abs/1811.12907} {arXiv:1811.12907 [astro-ph.HE]}
  \BibitemShut {NoStop}%
\bibitem [{\citenamefont {{Nitz}}\ \emph {et~al.}(2020)\citenamefont {{Nitz}},
  \citenamefont {{Dent}}, \citenamefont {{Davies}}, \citenamefont {{Kumar}},
  \citenamefont {{Capano}}, \citenamefont {{Harry}}, \citenamefont {{Mozzon}},
  \citenamefont {{Nuttall}}, \citenamefont {{Lundgren}},\ and\ \citenamefont
  {{T{\'a}pai}}}]{Nitz:2019hdf}%
  \BibitemOpen
  \bibfield  {author} {\bibinfo {author} {\bibfnamefont {A.~H.}\ \bibnamefont
  {{Nitz}}}, \bibinfo {author} {\bibfnamefont {T.}~\bibnamefont {{Dent}}},
  \bibinfo {author} {\bibfnamefont {G.~S.}\ \bibnamefont {{Davies}}}, \bibinfo
  {author} {\bibfnamefont {S.}~\bibnamefont {{Kumar}}}, \bibinfo {author}
  {\bibfnamefont {C.~D.}\ \bibnamefont {{Capano}}}, \bibinfo {author}
  {\bibfnamefont {I.}~\bibnamefont {{Harry}}}, \bibinfo {author} {\bibfnamefont
  {S.}~\bibnamefont {{Mozzon}}}, \bibinfo {author} {\bibfnamefont
  {L.}~\bibnamefont {{Nuttall}}}, \bibinfo {author} {\bibfnamefont
  {A.}~\bibnamefont {{Lundgren}}},\ and\ \bibinfo {author} {\bibfnamefont
  {M.}~\bibnamefont {{T{\'a}pai}}},\ }\bibfield  {title} {\bibinfo {title}
  {{2-OGC: Open Gravitational-wave Catalog of Binary Mergers from Analysis of
  Public Advanced LIGO and Virgo Data}},\ }\href
  {https://doi.org/10.3847/1538-4357/ab733f} {\bibfield  {journal} {\bibinfo
  {journal} {\apj}\ }\textbf {\bibinfo {volume} {891}},\ \bibinfo {eid} {123}
  (\bibinfo {year} {2020})},\ \Eprint {https://arxiv.org/abs/1910.05331}
  {arXiv:1910.05331 [astro-ph.HE]} \BibitemShut {NoStop}%
\bibitem [{\citenamefont {Abbott}\ \emph
  {et~al.}(2020{\natexlab{a}})\citenamefont {Abbott} \emph
  {et~al.}}]{Abbott:2020uma}%
  \BibitemOpen
  \bibfield  {author} {\bibinfo {author} {\bibfnamefont {B.}~\bibnamefont
  {Abbott}} \emph {et~al.} (\bibinfo {collaboration} {LIGO Scientific,
  Virgo}),\ }\bibfield  {title} {\bibinfo {title} {{GW190425: Observation of a
  Compact Binary Coalescence with Total Mass $\sim 3.4 M_{\odot}$}},\ }\href
  {https://doi.org/10.3847/2041-8213/ab75f5} {\bibfield  {journal} {\bibinfo
  {journal} {Astrophys. J. Lett.}\ }\textbf {\bibinfo {volume} {892}},\
  \bibinfo {pages} {L3} (\bibinfo {year} {2020}{\natexlab{a}})},\ \Eprint
  {https://arxiv.org/abs/2001.01761} {arXiv:2001.01761 [astro-ph.HE]}
  \BibitemShut {NoStop}%
\bibitem [{\citenamefont {Abbott}\ \emph
  {et~al.}(2020{\natexlab{b}})\citenamefont {Abbott} \emph
  {et~al.}}]{LIGOScientific:2020stg}%
  \BibitemOpen
  \bibfield  {author} {\bibinfo {author} {\bibfnamefont {R.}~\bibnamefont
  {Abbott}} \emph {et~al.} (\bibinfo {collaboration} {LIGO Scientific,
  Virgo}),\ }\bibfield  {title} {\bibinfo {title} {{GW190412: Observation of a
  Binary-Black-Hole Coalescence with Asymmetric Masses}},\ }\href@noop {} {\
  (\bibinfo {year} {2020}{\natexlab{b}})},\ \Eprint
  {https://arxiv.org/abs/2004.08342} {arXiv:2004.08342 [astro-ph.HE]}
  \BibitemShut {NoStop}%
\bibitem [{\citenamefont {Abbott}\ \emph
  {et~al.}(2020{\natexlab{c}})\citenamefont {Abbott} \emph
  {et~al.}}]{Abbott:2020khf}%
  \BibitemOpen
  \bibfield  {author} {\bibinfo {author} {\bibfnamefont {R.}~\bibnamefont
  {Abbott}} \emph {et~al.} (\bibinfo {collaboration} {LIGO Scientific,
  Virgo}),\ }\bibfield  {title} {\bibinfo {title} {{GW190814: Gravitational
  Waves from the Coalescence of a 23 Solar Mass Black Hole with a 2.6 Solar
  Mass Compact Object}},\ }\href {https://doi.org/10.3847/2041-8213/ab960f}
  {\bibfield  {journal} {\bibinfo  {journal} {Astrophys. J.}\ }\textbf
  {\bibinfo {volume} {896}},\ \bibinfo {pages} {L44} (\bibinfo {year}
  {2020}{\natexlab{c}})},\ \Eprint {https://arxiv.org/abs/2006.12611}
  {arXiv:2006.12611 [astro-ph.HE]} \BibitemShut {NoStop}%
\bibitem [{\citenamefont {Abbott}\ \emph
  {et~al.}(2020{\natexlab{d}})\citenamefont {Abbott} \emph
  {et~al.}}]{Abbott:2020niy}%
  \BibitemOpen
  \bibfield  {author} {\bibinfo {author} {\bibfnamefont {R.}~\bibnamefont
  {Abbott}} \emph {et~al.},\ }\bibfield  {title} {\bibinfo {title} {{GWTC-2:
  Compact Binary Coalescences Observed by LIGO and Virgo During the First Half
  of the Third Observing Run}},\ }\href@noop {} {\  (\bibinfo {year}
  {2020}{\natexlab{d}})},\ \Eprint {https://arxiv.org/abs/2010.14527}
  {arXiv:2010.14527 [gr-qc]} \BibitemShut {NoStop}%
\bibitem [{\citenamefont {Abbott}\ \emph
  {et~al.}(2020{\natexlab{e}})\citenamefont {Abbott} \emph
  {et~al.}}]{Abbott:2020tfl}%
  \BibitemOpen
  \bibfield  {author} {\bibinfo {author} {\bibfnamefont {R.}~\bibnamefont
  {Abbott}} \emph {et~al.} (\bibinfo {collaboration} {LIGO Scientific,
  Virgo}),\ }\bibfield  {title} {\bibinfo {title} {{GW190521: A Binary Black
  Hole Merger with a Total Mass of 150\,\,M\ensuremath{\odot}}},\ }\href
  {https://doi.org/10.1103/PhysRevLett.125.101102} {\bibfield  {journal}
  {\bibinfo  {journal} {Phys. Rev. Lett.}\ }\textbf {\bibinfo {volume} {125}},\
  \bibinfo {pages} {101102} (\bibinfo {year} {2020}{\natexlab{e}})},\ \Eprint
  {https://arxiv.org/abs/2009.01075} {arXiv:2009.01075 [gr-qc]} \BibitemShut
  {NoStop}%
\bibitem [{\citenamefont {{Armstrong}}\ \emph {et~al.}(1999)\citenamefont
  {{Armstrong}}, \citenamefont {{Estabrook}},\ and\ \citenamefont
  {{Tinto}}}]{1999ApJ...527..814A}%
  \BibitemOpen
  \bibfield  {author} {\bibinfo {author} {\bibfnamefont {J.~W.}\ \bibnamefont
  {{Armstrong}}}, \bibinfo {author} {\bibfnamefont {F.~B.}\ \bibnamefont
  {{Estabrook}}},\ and\ \bibinfo {author} {\bibfnamefont {M.}~\bibnamefont
  {{Tinto}}},\ }\bibfield  {title} {\bibinfo {title} {{Time-Delay
  Interferometry for Space-based Gravitational Wave Searches}},\ }\href
  {https://doi.org/10.1086/308110} {\bibfield  {journal} {\bibinfo  {journal}
  {\apj}\ }\textbf {\bibinfo {volume} {527}},\ \bibinfo {pages} {814} (\bibinfo
  {year} {1999})}\BibitemShut {NoStop}%
\bibitem [{\citenamefont {{Estabrook}}\ \emph {et~al.}(2000)\citenamefont
  {{Estabrook}}, \citenamefont {{Tinto}},\ and\ \citenamefont
  {{Armstrong}}}]{2000PhRvD..62d2002E}%
  \BibitemOpen
  \bibfield  {author} {\bibinfo {author} {\bibfnamefont {F.~B.}\ \bibnamefont
  {{Estabrook}}}, \bibinfo {author} {\bibfnamefont {M.}~\bibnamefont
  {{Tinto}}},\ and\ \bibinfo {author} {\bibfnamefont {J.~W.}\ \bibnamefont
  {{Armstrong}}},\ }\bibfield  {title} {\bibinfo {title} {{Time-delay analysis
  of LISA gravitational wave data: Elimination of spacecraft motion effects}},\
  }\href {https://doi.org/10.1103/PhysRevD.62.042002} {\bibfield  {journal}
  {\bibinfo  {journal} {\prd}\ }\textbf {\bibinfo {volume} {62}},\ \bibinfo
  {eid} {042002} (\bibinfo {year} {2000})}\BibitemShut {NoStop}%
\bibitem [{\citenamefont {{Armstrong}}\ \emph {et~al.}(2001)\citenamefont
  {{Armstrong}}, \citenamefont {{Estabrook}},\ and\ \citenamefont
  {{Tinto}}}]{2001CQGra..18.4059A}%
  \BibitemOpen
  \bibfield  {author} {\bibinfo {author} {\bibfnamefont {J.~W.}\ \bibnamefont
  {{Armstrong}}}, \bibinfo {author} {\bibfnamefont {F.~B.}\ \bibnamefont
  {{Estabrook}}},\ and\ \bibinfo {author} {\bibfnamefont {M.}~\bibnamefont
  {{Tinto}}},\ }\bibfield  {title} {\bibinfo {title} {{Sensitivities of
  alternate LISA configurations}},\ }\href
  {https://doi.org/10.1088/0264-9381/18/19/313} {\bibfield  {journal} {\bibinfo
   {journal} {Classical and Quantum Gravity}\ }\textbf {\bibinfo {volume}
  {18}},\ \bibinfo {pages} {4059} (\bibinfo {year} {2001})}\BibitemShut
  {NoStop}%
\bibitem [{\citenamefont {Larson}\ \emph {et~al.}(2002)\citenamefont {Larson},
  \citenamefont {Hellings},\ and\ \citenamefont {Hiscock}}]{Larson:2002xr}%
  \BibitemOpen
  \bibfield  {author} {\bibinfo {author} {\bibfnamefont {S.~L.}\ \bibnamefont
  {Larson}}, \bibinfo {author} {\bibfnamefont {R.~W.}\ \bibnamefont
  {Hellings}},\ and\ \bibinfo {author} {\bibfnamefont {W.~A.}\ \bibnamefont
  {Hiscock}},\ }\bibfield  {title} {\bibinfo {title} {{Unequal arm space borne
  gravitational wave detectors}},\ }\href
  {https://doi.org/10.1103/PhysRevD.66.062001} {\bibfield  {journal} {\bibinfo
  {journal} {Phys. Rev. D}\ }\textbf {\bibinfo {volume} {66}},\ \bibinfo
  {pages} {062001} (\bibinfo {year} {2002})},\ \Eprint
  {https://arxiv.org/abs/gr-qc/0206081} {arXiv:gr-qc/0206081} \BibitemShut
  {NoStop}%
\bibitem [{\citenamefont {Dhurandhar}\ \emph {et~al.}(2002)\citenamefont
  {Dhurandhar}, \citenamefont {Rajesh~Nayak},\ and\ \citenamefont
  {Vinet}}]{Dhurandhar:2002zcl}%
  \BibitemOpen
  \bibfield  {author} {\bibinfo {author} {\bibfnamefont {S.~V.}\ \bibnamefont
  {Dhurandhar}}, \bibinfo {author} {\bibfnamefont {K.}~\bibnamefont
  {Rajesh~Nayak}},\ and\ \bibinfo {author} {\bibfnamefont {J.~Y.}\ \bibnamefont
  {Vinet}},\ }\bibfield  {title} {\bibinfo {title} {{Algebraic approach to
  time-delay data analysis for LISA}},\ }\href
  {https://doi.org/10.1103/PhysRevD.65.102002} {\bibfield  {journal} {\bibinfo
  {journal} {Phys. Rev. D}\ }\textbf {\bibinfo {volume} {65}},\ \bibinfo
  {pages} {102002} (\bibinfo {year} {2002})},\ \Eprint
  {https://arxiv.org/abs/gr-qc/0112059} {arXiv:gr-qc/0112059 [gr-qc]}
  \BibitemShut {NoStop}%
\bibitem [{\citenamefont {{Tinto}}\ \emph {et~al.}(2003)\citenamefont
  {{Tinto}}, \citenamefont {{Shaddock}}, \citenamefont {{Sylvestre}},\ and\
  \citenamefont {{Armstrong}}}]{2003PhRvD..67l2003T}%
  \BibitemOpen
  \bibfield  {author} {\bibinfo {author} {\bibfnamefont {M.}~\bibnamefont
  {{Tinto}}}, \bibinfo {author} {\bibfnamefont {D.~A.}\ \bibnamefont
  {{Shaddock}}}, \bibinfo {author} {\bibfnamefont {J.}~\bibnamefont
  {{Sylvestre}}},\ and\ \bibinfo {author} {\bibfnamefont {J.~W.}\ \bibnamefont
  {{Armstrong}}},\ }\bibfield  {title} {\bibinfo {title} {{Implementation of
  time-delay interferometry for LISA}},\ }\href
  {https://doi.org/10.1103/PhysRevD.67.122003} {\bibfield  {journal} {\bibinfo
  {journal} {\prd}\ }\textbf {\bibinfo {volume} {67}},\ \bibinfo {eid} {122003}
  (\bibinfo {year} {2003})},\ \Eprint {https://arxiv.org/abs/gr-qc/0303013}
  {arXiv:gr-qc/0303013 [gr-qc]} \BibitemShut {NoStop}%
\bibitem [{\citenamefont {Vallisneri}(2005{\natexlab{a}})}]{Vallisneri:2004bn}%
  \BibitemOpen
  \bibfield  {author} {\bibinfo {author} {\bibfnamefont {M.}~\bibnamefont
  {Vallisneri}},\ }\bibfield  {title} {\bibinfo {title} {{Synthetic LISA:
  Simulating time delay interferometry in a model LISA}},\ }\href
  {https://doi.org/10.1103/PhysRevD.71.022001} {\bibfield  {journal} {\bibinfo
  {journal} {Phys. Rev. D}\ }\textbf {\bibinfo {volume} {71}},\ \bibinfo
  {pages} {022001} (\bibinfo {year} {2005}{\natexlab{a}})},\ \Eprint
  {https://arxiv.org/abs/gr-qc/0407102} {arXiv:gr-qc/0407102 [gr-qc]}
  \BibitemShut {NoStop}%
\bibitem [{\citenamefont {{Petiteau}}\ \emph {et~al.}(2008)\citenamefont
  {{Petiteau}}, \citenamefont {{Auger}}, \citenamefont {{Halloin}},
  \citenamefont {{Jeannin}}, \citenamefont {{Plagnol}}, \citenamefont
  {{Pireaux}}, \citenamefont {{Regimbau}},\ and\ \citenamefont
  {{Vinet}}}]{2008PhRvD..77b3002P}%
  \BibitemOpen
  \bibfield  {author} {\bibinfo {author} {\bibfnamefont {A.}~\bibnamefont
  {{Petiteau}}}, \bibinfo {author} {\bibfnamefont {G.}~\bibnamefont {{Auger}}},
  \bibinfo {author} {\bibfnamefont {H.}~\bibnamefont {{Halloin}}}, \bibinfo
  {author} {\bibfnamefont {O.}~\bibnamefont {{Jeannin}}}, \bibinfo {author}
  {\bibfnamefont {E.}~\bibnamefont {{Plagnol}}}, \bibinfo {author}
  {\bibfnamefont {S.}~\bibnamefont {{Pireaux}}}, \bibinfo {author}
  {\bibfnamefont {T.}~\bibnamefont {{Regimbau}}},\ and\ \bibinfo {author}
  {\bibfnamefont {J.-Y.}\ \bibnamefont {{Vinet}}},\ }\bibfield  {title}
  {\bibinfo {title} {{LISACode: A scientific simulator of LISA}},\ }\href
  {https://doi.org/10.1103/PhysRevD.77.023002} {\bibfield  {journal} {\bibinfo
  {journal} {\prd}\ }\textbf {\bibinfo {volume} {77}},\ \bibinfo {eid} {023002}
  (\bibinfo {year} {2008})},\ \Eprint {https://arxiv.org/abs/0802.2023}
  {arXiv:0802.2023 [gr-qc]} \BibitemShut {NoStop}%
\bibitem [{\citenamefont {Tinto}\ and\ \citenamefont
  {Dhurandhar}(2021)}]{Tinto:2020fcc}%
  \BibitemOpen
  \bibfield  {author} {\bibinfo {author} {\bibfnamefont {M.}~\bibnamefont
  {Tinto}}\ and\ \bibinfo {author} {\bibfnamefont {S.~V.}\ \bibnamefont
  {Dhurandhar}},\ }\bibfield  {title} {\bibinfo {title} {{Time-delay
  interferometry}},\ }\href {https://doi.org/10.1007/s41114-020-00029-6}
  {\bibfield  {journal} {\bibinfo  {journal} {Living Rev. Rel.}\ }\textbf
  {\bibinfo {volume} {24}},\ \bibinfo {pages} {1} (\bibinfo {year}
  {2021})}\BibitemShut {NoStop}%
\bibitem [{\citenamefont {Shaddock}\ \emph {et~al.}(2003)\citenamefont
  {Shaddock}, \citenamefont {Tinto}, \citenamefont {Estabrook},\ and\
  \citenamefont {Armstrong}}]{Shaddock:2003dj}%
  \BibitemOpen
  \bibfield  {author} {\bibinfo {author} {\bibfnamefont {D.~A.}\ \bibnamefont
  {Shaddock}}, \bibinfo {author} {\bibfnamefont {M.}~\bibnamefont {Tinto}},
  \bibinfo {author} {\bibfnamefont {F.~B.}\ \bibnamefont {Estabrook}},\ and\
  \bibinfo {author} {\bibfnamefont {J.}~\bibnamefont {Armstrong}},\ }\bibfield
  {title} {\bibinfo {title} {{Data combinations accounting for LISA spacecraft
  motion}},\ }\href {https://doi.org/10.1103/PhysRevD.68.061303} {\bibfield
  {journal} {\bibinfo  {journal} {Phys. Rev. D}\ }\textbf {\bibinfo {volume}
  {68}},\ \bibinfo {pages} {061303} (\bibinfo {year} {2003})},\ \Eprint
  {https://arxiv.org/abs/gr-qc/0307080} {arXiv:gr-qc/0307080} \BibitemShut
  {NoStop}%
\bibitem [{\citenamefont {Cornish}\ and\ \citenamefont
  {Hellings}(2003)}]{Cornish:2003tz}%
  \BibitemOpen
  \bibfield  {author} {\bibinfo {author} {\bibfnamefont {N.~J.}\ \bibnamefont
  {Cornish}}\ and\ \bibinfo {author} {\bibfnamefont {R.~W.}\ \bibnamefont
  {Hellings}},\ }\bibfield  {title} {\bibinfo {title} {{The Effects of orbital
  motion on LISA time delay interferometry}},\ }\href
  {https://doi.org/10.1088/0264-9381/20/22/009} {\bibfield  {journal} {\bibinfo
   {journal} {Class. Quant. Grav.}\ }\textbf {\bibinfo {volume} {20}},\
  \bibinfo {pages} {4851} (\bibinfo {year} {2003})},\ \Eprint
  {https://arxiv.org/abs/gr-qc/0306096} {arXiv:gr-qc/0306096 [gr-qc]}
  \BibitemShut {NoStop}%
\bibitem [{\citenamefont {Tinto}\ \emph {et~al.}(2004)\citenamefont {Tinto},
  \citenamefont {Estabrook},\ and\ \citenamefont {Armstrong}}]{Tinto:2003vj}%
  \BibitemOpen
  \bibfield  {author} {\bibinfo {author} {\bibfnamefont {M.}~\bibnamefont
  {Tinto}}, \bibinfo {author} {\bibfnamefont {F.~B.}\ \bibnamefont
  {Estabrook}},\ and\ \bibinfo {author} {\bibfnamefont {J.}~\bibnamefont
  {Armstrong}},\ }\bibfield  {title} {\bibinfo {title} {{Time delay
  interferometry with moving spacecraft arrays}},\ }\href
  {https://doi.org/10.1103/PhysRevD.69.082001} {\bibfield  {journal} {\bibinfo
  {journal} {Phys. Rev. D}\ }\textbf {\bibinfo {volume} {69}},\ \bibinfo
  {pages} {082001} (\bibinfo {year} {2004})},\ \Eprint
  {https://arxiv.org/abs/gr-qc/0310017} {arXiv:gr-qc/0310017} \BibitemShut
  {NoStop}%
\bibitem [{\citenamefont {Dhurandhar}\ \emph {et~al.}(2010)\citenamefont
  {Dhurandhar}, \citenamefont {Nayak},\ and\ \citenamefont
  {Vinet}}]{Dhurandhar:2010pd}%
  \BibitemOpen
  \bibfield  {author} {\bibinfo {author} {\bibfnamefont {S.}~\bibnamefont
  {Dhurandhar}}, \bibinfo {author} {\bibfnamefont {K.}~\bibnamefont {Nayak}},\
  and\ \bibinfo {author} {\bibfnamefont {J.}~\bibnamefont {Vinet}},\ }\bibfield
   {title} {\bibinfo {title} {{Time Delay Interferometry for LISA with one arm
  dysfunctional}},\ }\href {https://doi.org/10.1088/0264-9381/27/13/135013}
  {\bibfield  {journal} {\bibinfo  {journal} {Class. Quant. Grav.}\ }\textbf
  {\bibinfo {volume} {27}},\ \bibinfo {pages} {135013} (\bibinfo {year}
  {2010})},\ \Eprint {https://arxiv.org/abs/1001.4911} {arXiv:1001.4911
  [gr-qc]} \BibitemShut {NoStop}%
\bibitem [{\citenamefont {Tinto}\ and\ \citenamefont
  {Hartwig}(2018)}]{Tinto:2018kij}%
  \BibitemOpen
  \bibfield  {author} {\bibinfo {author} {\bibfnamefont {M.}~\bibnamefont
  {Tinto}}\ and\ \bibinfo {author} {\bibfnamefont {O.}~\bibnamefont
  {Hartwig}},\ }\bibfield  {title} {\bibinfo {title} {{Time-Delay
  Interferometry and Clock-Noise Calibration}},\ }\href
  {https://doi.org/10.1103/PhysRevD.98.042003} {\bibfield  {journal} {\bibinfo
  {journal} {Phys. Rev. D}\ }\textbf {\bibinfo {volume} {98}},\ \bibinfo
  {pages} {042003} (\bibinfo {year} {2018})},\ \Eprint
  {https://arxiv.org/abs/1807.02594} {arXiv:1807.02594 [gr-qc]} \BibitemShut
  {NoStop}%
\bibitem [{\citenamefont {{Bayle}}\ \emph {et~al.}(2019)\citenamefont
  {{Bayle}}, \citenamefont {{Lilley}}, \citenamefont {{Petiteau}},\ and\
  \citenamefont {{Halloin}}}]{2019PhRvD..99h4023B}%
  \BibitemOpen
  \bibfield  {author} {\bibinfo {author} {\bibfnamefont {J.-B.}\ \bibnamefont
  {{Bayle}}}, \bibinfo {author} {\bibfnamefont {M.}~\bibnamefont {{Lilley}}},
  \bibinfo {author} {\bibfnamefont {A.}~\bibnamefont {{Petiteau}}},\ and\
  \bibinfo {author} {\bibfnamefont {H.}~\bibnamefont {{Halloin}}},\ }\bibfield
  {title} {\bibinfo {title} {{Effect of filters on the time-delay
  interferometry residual laser noise for LISA}},\ }\href
  {https://doi.org/10.1103/PhysRevD.99.084023} {\bibfield  {journal} {\bibinfo
  {journal} {\prd}\ }\textbf {\bibinfo {volume} {99}},\ \bibinfo {eid} {084023}
  (\bibinfo {year} {2019})},\ \Eprint {https://arxiv.org/abs/1811.01575}
  {arXiv:1811.01575 [astro-ph.IM]} \BibitemShut {NoStop}%
\bibitem [{\citenamefont {{Muratore}}\ \emph {et~al.}(2020)\citenamefont
  {{Muratore}}, \citenamefont {{Vetrugno}},\ and\ \citenamefont
  {{Vitale}}}]{2020arXiv200111221M}%
  \BibitemOpen
  \bibfield  {author} {\bibinfo {author} {\bibfnamefont {M.}~\bibnamefont
  {{Muratore}}}, \bibinfo {author} {\bibfnamefont {D.}~\bibnamefont
  {{Vetrugno}}},\ and\ \bibinfo {author} {\bibfnamefont {S.}~\bibnamefont
  {{Vitale}}},\ }\bibfield  {title} {\bibinfo {title} {{Revisitation of time
  delay interferometry combinations that suppress laser noise in LISA}},\
  }\href@noop {} {\bibfield  {journal} {\bibinfo  {journal} {arXiv e-prints}\ }
  (\bibinfo {year} {2020})},\ \Eprint {https://arxiv.org/abs/2001.11221}
  {arXiv:2001.11221 [astro-ph.IM]} \BibitemShut {NoStop}%
\bibitem [{\citenamefont {Vallisneri}\ \emph {et~al.}(2020)\citenamefont
  {Vallisneri}, \citenamefont {Bayle}, \citenamefont {Babak},\ and\
  \citenamefont {Petiteau}}]{Vallisneri:2020otf}%
  \BibitemOpen
  \bibfield  {author} {\bibinfo {author} {\bibfnamefont {M.}~\bibnamefont
  {Vallisneri}}, \bibinfo {author} {\bibfnamefont {J.-B.}\ \bibnamefont
  {Bayle}}, \bibinfo {author} {\bibfnamefont {S.}~\bibnamefont {Babak}},\ and\
  \bibinfo {author} {\bibfnamefont {A.}~\bibnamefont {Petiteau}},\ }\bibfield
  {title} {\bibinfo {title} {{TDI-infinity: time-delay interferometry without
  delays}},\ }\href@noop {} {\  (\bibinfo {year} {2020})},\ \Eprint
  {https://arxiv.org/abs/2008.12343} {arXiv:2008.12343 [gr-qc]} \BibitemShut
  {NoStop}%
\bibitem [{\citenamefont {Otto}\ \emph {et~al.}(2012)\citenamefont {Otto},
  \citenamefont {Heinzel},\ and\ \citenamefont {Danzmann}}]{Otto:2012dk}%
  \BibitemOpen
  \bibfield  {author} {\bibinfo {author} {\bibfnamefont {M.}~\bibnamefont
  {Otto}}, \bibinfo {author} {\bibfnamefont {G.}~\bibnamefont {Heinzel}},\ and\
  \bibinfo {author} {\bibfnamefont {K.}~\bibnamefont {Danzmann}},\ }\bibfield
  {title} {\bibinfo {title} {{TDI and clock noise removal for the split
  interferometry configuration of LISA}},\ }\href
  {https://doi.org/10.1088/0264-9381/29/20/205003} {\bibfield  {journal}
  {\bibinfo  {journal} {Class. Quant. Grav.}\ }\textbf {\bibinfo {volume}
  {29}},\ \bibinfo {pages} {205003} (\bibinfo {year} {2012})}\BibitemShut
  {NoStop}%
\bibitem [{\citenamefont {Otto}(2015)}]{Otto:2015}%
  \BibitemOpen
  \bibfield  {author} {\bibinfo {author} {\bibfnamefont {M.}~\bibnamefont
  {Otto}},\ }\href@noop {} {\bibinfo {title} {{Time-Delay Interferometry
  Simulations for the Laser Interferometer Space Antenna}}} (\bibinfo {year}
  {2015})\BibitemShut {NoStop}%
\bibitem [{\citenamefont {Hartwig}\ and\ \citenamefont
  {Bayle}(2020)}]{Hartwig:2020tdu}%
  \BibitemOpen
  \bibfield  {author} {\bibinfo {author} {\bibfnamefont {O.}~\bibnamefont
  {Hartwig}}\ and\ \bibinfo {author} {\bibfnamefont {J.-B.}\ \bibnamefont
  {Bayle}},\ }\bibfield  {title} {\bibinfo {title} {{Clock-jitter reduction in
  LISA time-delay interferometry combinations}},\ }\href@noop {} {\  (\bibinfo
  {year} {2020})},\ \Eprint {https://arxiv.org/abs/2005.02430}
  {arXiv:2005.02430 [astro-ph.IM]} \BibitemShut {NoStop}%
\bibitem [{\citenamefont {Chwalla}\ \emph {et~al.}(2016)\citenamefont {Chwalla}
  \emph {et~al.}}]{Chwalla:2016bzk}%
  \BibitemOpen
  \bibfield  {author} {\bibinfo {author} {\bibfnamefont {M.}~\bibnamefont
  {Chwalla}} \emph {et~al.},\ }\bibfield  {title} {\bibinfo {title} {{Design
  and construction of an optical test bed for LISA imaging systems and
  tilt-to-length coupling}},\ }\href
  {https://doi.org/10.1088/0264-9381/33/24/245015} {\bibfield  {journal}
  {\bibinfo  {journal} {Class. Quant. Grav.}\ }\textbf {\bibinfo {volume}
  {33}},\ \bibinfo {pages} {245015} (\bibinfo {year} {2016})},\ \Eprint
  {https://arxiv.org/abs/1607.00408} {arXiv:1607.00408 [astro-ph.IM]}
  \BibitemShut {NoStop}%
\bibitem [{\citenamefont {Tr\"obs}\ \emph {et~al.}(2018)\citenamefont {Tr\"obs}
  \emph {et~al.}}]{Trobs:2017msu}%
  \BibitemOpen
  \bibfield  {author} {\bibinfo {author} {\bibfnamefont {M.}~\bibnamefont
  {Tr\"obs}} \emph {et~al.},\ }\bibfield  {title} {\bibinfo {title} {{Reducing
  tilt-to-length coupling for the LISA test mass interferometer}},\ }\href
  {https://doi.org/10.1088/1361-6382/aab86c} {\bibfield  {journal} {\bibinfo
  {journal} {Class. Quant. Grav.}\ }\textbf {\bibinfo {volume} {35}},\ \bibinfo
  {pages} {105001} (\bibinfo {year} {2018})},\ \Eprint
  {https://arxiv.org/abs/1711.10320} {arXiv:1711.10320 [astro-ph.IM]}
  \BibitemShut {NoStop}%
\bibitem [{\citenamefont {{Cornish}}\ and\ \citenamefont
  {{Rubbo}}(2003)}]{2003PhRvD..67b2001C}%
  \BibitemOpen
  \bibfield  {author} {\bibinfo {author} {\bibfnamefont {N.~J.}\ \bibnamefont
  {{Cornish}}}\ and\ \bibinfo {author} {\bibfnamefont {L.~J.}\ \bibnamefont
  {{Rubbo}}},\ }\bibfield  {title} {\bibinfo {title} {{LISA response
  function}},\ }\href {https://doi.org/10.1103/PhysRevD.67.022001} {\bibfield
  {journal} {\bibinfo  {journal} {\prd}\ }\textbf {\bibinfo {volume} {67}},\
  \bibinfo {eid} {022001} (\bibinfo {year} {2003})}\BibitemShut {NoStop}%
\bibitem [{\citenamefont {{Rubbo}}\ \emph {et~al.}(2004)\citenamefont
  {{Rubbo}}, \citenamefont {{Cornish}},\ and\ \citenamefont
  {{Poujade}}}]{2004PhRvD..69h2003R}%
  \BibitemOpen
  \bibfield  {author} {\bibinfo {author} {\bibfnamefont {L.~J.}\ \bibnamefont
  {{Rubbo}}}, \bibinfo {author} {\bibfnamefont {N.~J.}\ \bibnamefont
  {{Cornish}}},\ and\ \bibinfo {author} {\bibfnamefont {O.}~\bibnamefont
  {{Poujade}}},\ }\bibfield  {title} {\bibinfo {title} {{Forward modeling of
  space-borne gravitational wave detectors}},\ }\href
  {https://doi.org/10.1103/PhysRevD.69.082003} {\bibfield  {journal} {\bibinfo
  {journal} {\prd}\ }\textbf {\bibinfo {volume} {69}},\ \bibinfo {eid} {082003}
  (\bibinfo {year} {2004})},\ \Eprint {https://arxiv.org/abs/gr-qc/0311069}
  {arXiv:gr-qc/0311069 [gr-qc]} \BibitemShut {NoStop}%
\bibitem [{\citenamefont {Ni}(2013)}]{Ni:2013}%
  \BibitemOpen
  \bibfield  {author} {\bibinfo {author} {\bibfnamefont {W.-T.}\ \bibnamefont
  {Ni}},\ }\bibfield  {title} {\bibinfo {title} {{ASTROD-GW: Overview and
  Progress}},\ }\href {https://doi.org/10.1142/S0218271813410046} {\bibfield
  {journal} {\bibinfo  {journal} {Int. J. Mod. Phys. D}\ }\textbf {\bibinfo
  {volume} {22}},\ \bibinfo {pages} {1341004} (\bibinfo {year} {2013})},\
  \Eprint {https://arxiv.org/abs/1212.2816} {arXiv:1212.2816 [astro-ph.IM]}
  \BibitemShut {NoStop}%
\bibitem [{\citenamefont {Wang}(2011)}]{Wang:2011}%
  \BibitemOpen
  \bibfield  {author} {\bibinfo {author} {\bibfnamefont {G.}~\bibnamefont
  {Wang}},\ }\href@noop {} {\bibinfo {title} {{Time-delay Interferometry for
  ASTROD-GW}}} (\bibinfo {year} {2011})\BibitemShut {NoStop}%
\bibitem [{\citenamefont {Wang}\ and\ \citenamefont {Ni}(2012)}]{Wang:2014aea}%
  \BibitemOpen
  \bibfield  {author} {\bibinfo {author} {\bibfnamefont {G.}~\bibnamefont
  {Wang}}\ and\ \bibinfo {author} {\bibfnamefont {W.-T.}\ \bibnamefont {Ni}},\
  }\bibfield  {title} {\bibinfo {title} {{Time-delay Interferometry for
  ASTROD-GW}},\ }\href {https://doi.org/10.1016/j.chinastron.2012.04.009}
  {\bibfield  {journal} {\bibinfo  {journal} {Chin. Astron. Astrophys.}\
  }\textbf {\bibinfo {volume} {36}},\ \bibinfo {pages} {211} (\bibinfo {year}
  {2012})},\ \bibinfo {note} {and references therein}\BibitemShut {NoStop}%
\bibitem [{\citenamefont {Wang}\ and\ \citenamefont
  {Ni}(2013{\natexlab{a}})}]{Wang:2012ce}%
  \BibitemOpen
  \bibfield  {author} {\bibinfo {author} {\bibfnamefont {G.}~\bibnamefont
  {Wang}}\ and\ \bibinfo {author} {\bibfnamefont {W.-T.}\ \bibnamefont {Ni}},\
  }\bibfield  {title} {\bibinfo {title} {{Numermcal simulation of time delay
  interferometry for NGO/eLISA}},\ }\href
  {https://doi.org/10.1088/0264-9381/30/6/065011} {\bibfield  {journal}
  {\bibinfo  {journal} {Class. Quant. Grav.}\ }\textbf {\bibinfo {volume}
  {30}},\ \bibinfo {pages} {065011} (\bibinfo {year} {2013}{\natexlab{a}})},\
  \Eprint {https://arxiv.org/abs/1204.2125} {arXiv:1204.2125 [gr-qc]}
  \BibitemShut {NoStop}%
\bibitem [{\citenamefont {Wang}\ and\ \citenamefont
  {Ni}(2013{\natexlab{b}})}]{Wang:2012te}%
  \BibitemOpen
  \bibfield  {author} {\bibinfo {author} {\bibfnamefont {G.}~\bibnamefont
  {Wang}}\ and\ \bibinfo {author} {\bibfnamefont {W.-T.}\ \bibnamefont {Ni}},\
  }\bibfield  {title} {\bibinfo {title} {{Orbit optimization for ASTROD-GW and
  its time delay interferometry with two arms using CGC ephemeris}},\ }\href
  {https://doi.org/10.1088/1674-1056/22/4/049501} {\bibfield  {journal}
  {\bibinfo  {journal} {Chin. Phys.}\ }\textbf {\bibinfo {volume} {B22}},\
  \bibinfo {pages} {049501} (\bibinfo {year} {2013}{\natexlab{b}})},\ \Eprint
  {https://arxiv.org/abs/1205.5175} {arXiv:1205.5175 [gr-qc]} \BibitemShut
  {NoStop}%
\bibitem [{\citenamefont {Dhurandhar}\ \emph {et~al.}(2013)\citenamefont
  {Dhurandhar}, \citenamefont {Ni},\ and\ \citenamefont
  {Wang}}]{Dhurandhar:2011ik}%
  \BibitemOpen
  \bibfield  {author} {\bibinfo {author} {\bibfnamefont {S.~V.}\ \bibnamefont
  {Dhurandhar}}, \bibinfo {author} {\bibfnamefont {W.~T.}\ \bibnamefont {Ni}},\
  and\ \bibinfo {author} {\bibfnamefont {G.}~\bibnamefont {Wang}},\ }\bibfield
  {title} {\bibinfo {title} {{Numerical simulation of time delay interferometry
  for a LISA-like mission with the simplification of having only one
  interferometer}},\ }\href {https://doi.org/10.1016/j.asr.2012.09.017}
  {\bibfield  {journal} {\bibinfo  {journal} {Adv. Space Res.}\ }\textbf
  {\bibinfo {volume} {51}},\ \bibinfo {pages} {198} (\bibinfo {year} {2013})},\
  \Eprint {https://arxiv.org/abs/1102.4965} {arXiv:1102.4965 [gr-qc]}
  \BibitemShut {NoStop}%
\bibitem [{\citenamefont {Wang}\ and\ \citenamefont {Ni}(2015)}]{Wang:2014cla}%
  \BibitemOpen
  \bibfield  {author} {\bibinfo {author} {\bibfnamefont {G.}~\bibnamefont
  {Wang}}\ and\ \bibinfo {author} {\bibfnamefont {W.-T.}\ \bibnamefont {Ni}},\
  }\bibfield  {title} {\bibinfo {title} {{Orbit optimization and time delay
  interferometry for inclined ASTROD-GW formation with half-year
  precession-period}},\ }\href {https://doi.org/10.1088/1674-1056/24/5/059501}
  {\bibfield  {journal} {\bibinfo  {journal} {Chin. Phys.}\ }\textbf {\bibinfo
  {volume} {B24}},\ \bibinfo {pages} {059501} (\bibinfo {year} {2015})},\
  \Eprint {https://arxiv.org/abs/1409.4162} {arXiv:1409.4162 [gr-qc]}
  \BibitemShut {NoStop}%
\bibitem [{\citenamefont {Wang}\ and\ \citenamefont {Ni}(2019)}]{Wang:2017aqq}%
  \BibitemOpen
  \bibfield  {author} {\bibinfo {author} {\bibfnamefont {G.}~\bibnamefont
  {Wang}}\ and\ \bibinfo {author} {\bibfnamefont {W.-T.}\ \bibnamefont {Ni}},\
  }\bibfield  {title} {\bibinfo {title} {{Numerical simulation of time delay
  interferometry for TAIJI and new LISA}},\ }\href
  {https://doi.org/10.1088/1674-4527/19/4/58} {\bibfield  {journal} {\bibinfo
  {journal} {Res. Astron. Astrophys.}\ }\textbf {\bibinfo {volume} {19}},\
  \bibinfo {pages} {058} (\bibinfo {year} {2019})},\ \Eprint
  {https://arxiv.org/abs/1707.09127} {arXiv:1707.09127 [astro-ph.IM]}
  \BibitemShut {NoStop}%
\bibitem [{\citenamefont {Wang}\ \emph
  {et~al.}(2020{\natexlab{a}})\citenamefont {Wang}, \citenamefont {Ni},\ and\
  \citenamefont {Wu}}]{Wang:2019ipi}%
  \BibitemOpen
  \bibfield  {author} {\bibinfo {author} {\bibfnamefont {G.}~\bibnamefont
  {Wang}}, \bibinfo {author} {\bibfnamefont {W.-T.}\ \bibnamefont {Ni}},\ and\
  \bibinfo {author} {\bibfnamefont {A.-M.}\ \bibnamefont {Wu}},\ }\bibfield
  {title} {\bibinfo {title} {{Orbit design and thruster requirement for various
  constant-arm space mission concepts for gravitational-wave observation}},\
  }\href {https://doi.org/10.1142/S0218271819400066} {\bibfield  {journal}
  {\bibinfo  {journal} {Int. J. Mod. Phys. D}\ }\textbf {\bibinfo {volume}
  {29}},\ \bibinfo {pages} {1940006} (\bibinfo {year} {2020}{\natexlab{a}})},\
  \Eprint {https://arxiv.org/abs/1908.05444} {arXiv:1908.05444 [gr-qc]}
  \BibitemShut {NoStop}%
\bibitem [{\citenamefont {Wang}\ \emph
  {et~al.}(2020{\natexlab{b}})\citenamefont {Wang}, \citenamefont {Ni},\ and\
  \citenamefont {Han}}]{Wang:2020fwa}%
  \BibitemOpen
  \bibfield  {author} {\bibinfo {author} {\bibfnamefont {G.}~\bibnamefont
  {Wang}}, \bibinfo {author} {\bibfnamefont {W.-T.}\ \bibnamefont {Ni}},\ and\
  \bibinfo {author} {\bibfnamefont {W.-B.}\ \bibnamefont {Han}},\ }\bibfield
  {title} {\bibinfo {title} {{Revisiting time delay interferometry for
  unequal-arm LISA and TAIJI}},\ }\href@noop {} {\  (\bibinfo {year}
  {2020}{\natexlab{b}})},\ \bibinfo {note} {(Paper I)},\ \Eprint
  {https://arxiv.org/abs/2008.05812} {arXiv:2008.05812 [gr-qc]} \BibitemShut
  {NoStop}%
\bibitem [{\citenamefont {Vallisneri}(2005{\natexlab{b}})}]{Vallisneri:2005ji}%
  \BibitemOpen
  \bibfield  {author} {\bibinfo {author} {\bibfnamefont {M.}~\bibnamefont
  {Vallisneri}},\ }\bibfield  {title} {\bibinfo {title} {{Geometric time delay
  interferometry}},\ }\href {https://doi.org/10.1103/PhysRevD.76.109903,
  10.1103/PhysRevD.72.042003} {\bibfield  {journal} {\bibinfo  {journal} {Phys.
  Rev. D}\ }\textbf {\bibinfo {volume} {72}},\ \bibinfo {pages} {042003}
  (\bibinfo {year} {2005}{\natexlab{b}})},\ \bibinfo {note} {[Erratum: Phys.
  Rev. D 76, 109903(2007)]},\ \Eprint {https://arxiv.org/abs/gr-qc/0504145}
  {arXiv:gr-qc/0504145 [gr-qc]} \BibitemShut {NoStop}%
\bibitem [{\citenamefont {{Wang}}\ \emph {et~al.}(2020)\citenamefont {{Wang}},
  \citenamefont {{Ni}}, \citenamefont {{Han}}, \citenamefont {{Yang}},\ and\
  \citenamefont {{Zhong}}}]{Wang:2020a}%
  \BibitemOpen
  \bibfield  {author} {\bibinfo {author} {\bibfnamefont {G.}~\bibnamefont
  {{Wang}}}, \bibinfo {author} {\bibfnamefont {W.-T.}\ \bibnamefont {{Ni}}},
  \bibinfo {author} {\bibfnamefont {W.-B.}\ \bibnamefont {{Han}}}, \bibinfo
  {author} {\bibfnamefont {S.-C.}\ \bibnamefont {{Yang}}},\ and\ \bibinfo
  {author} {\bibfnamefont {X.-Y.}\ \bibnamefont {{Zhong}}},\ }\bibfield
  {title} {\bibinfo {title} {{Numerical simulation of sky localization for
  LISA-TAIJI joint observation}},\ }\href
  {https://doi.org/10.1103/PhysRevD.102.024089} {\bibfield  {journal} {\bibinfo
   {journal} {\prd}\ }\textbf {\bibinfo {volume} {102}},\ \bibinfo {pages}
  {024089} (\bibinfo {year} {2020})},\ \Eprint
  {https://arxiv.org/abs/2002.12628} {arXiv:2002.12628} \BibitemShut {NoStop}%
\bibitem [{\citenamefont {{Amaro-Seoane}}\ \emph {et~al.}(2017)\citenamefont
  {{Amaro-Seoane}}, \citenamefont {{Audley}}, \citenamefont {{Babak}},\ and\
  \citenamefont {{et al}}}]{2017arXiv170200786A}%
  \BibitemOpen
  \bibfield  {author} {\bibinfo {author} {\bibfnamefont {P.}~\bibnamefont
  {{Amaro-Seoane}}}, \bibinfo {author} {\bibfnamefont {H.}~\bibnamefont
  {{Audley}}}, \bibinfo {author} {\bibfnamefont {S.}~\bibnamefont {{Babak}}},\
  and\ \bibinfo {author} {\bibnamefont {{et al}}} (\bibinfo {collaboration}
  {{LISA Team}}),\ }\bibfield  {title} {\bibinfo {title} {{Laser Interferometer
  Space Antenna}},\ }\href@noop {} {\bibfield  {journal} {\bibinfo  {journal}
  {arXiv e-prints}\ ,\ \bibinfo {eid} {arXiv:1702.00786}} (\bibinfo {year}
  {2017})}\BibitemShut {NoStop}%
\bibitem [{\citenamefont {Ashby}\ and\ \citenamefont
  {Bender}(2008)}]{Ashby:2008lea}%
  \BibitemOpen
  \bibfield  {author} {\bibinfo {author} {\bibfnamefont {N.}~\bibnamefont
  {Ashby}}\ and\ \bibinfo {author} {\bibfnamefont {P.~L.}\ \bibnamefont
  {Bender}},\ }\bibfield  {title} {\bibinfo {title} {{Measurement of the
  Shapiro Time Delay Between Drag-Free Spacecraft}},\ }\href
  {https://doi.org/10.1007/978-3-540-34377-6_10} {\bibfield  {journal}
  {\bibinfo  {journal} {Astrophys. Space Sci. Libr.}\ }\textbf {\bibinfo
  {volume} {349}},\ \bibinfo {pages} {219} (\bibinfo {year}
  {2008})}\BibitemShut {NoStop}%
\bibitem [{\citenamefont {Shapiro}(1964)}]{Shapiro:1964uw}%
  \BibitemOpen
  \bibfield  {author} {\bibinfo {author} {\bibfnamefont {I.~I.}\ \bibnamefont
  {Shapiro}},\ }\bibfield  {title} {\bibinfo {title} {{Fourth Test of General
  Relativity}},\ }\href {https://doi.org/10.1103/PhysRevLett.13.789} {\bibfield
   {journal} {\bibinfo  {journal} {Phys. Rev. Lett.}\ }\textbf {\bibinfo
  {volume} {13}},\ \bibinfo {pages} {789} (\bibinfo {year} {1964})}\BibitemShut
  {NoStop}%
\bibitem [{\citenamefont {Kopeikin}(2009)}]{Kopeikin:2008xv}%
  \BibitemOpen
  \bibfield  {author} {\bibinfo {author} {\bibfnamefont {S.~M.}\ \bibnamefont
  {Kopeikin}},\ }\bibfield  {title} {\bibinfo {title} {{Post-Newtonian
  limitations on measurement of the PPN parameters caused by motion of
  gravitating bodies}},\ }\href
  {https://doi.org/10.1111/j.1365-2966.2009.15387.x} {\bibfield  {journal}
  {\bibinfo  {journal} {Mon. Not. Roy. Astron. Soc.}\ }\textbf {\bibinfo
  {volume} {399}},\ \bibinfo {pages} {1539} (\bibinfo {year} {2009})},\ \Eprint
  {https://arxiv.org/abs/0809.3433} {arXiv:0809.3433 [gr-qc]} \BibitemShut
  {NoStop}%
\bibitem [{\citenamefont {{Newhall}}(1989)}]{Newhall:1989CeMec}%
  \BibitemOpen
  \bibfield  {author} {\bibinfo {author} {\bibfnamefont {X.~X.}\ \bibnamefont
  {{Newhall}}},\ }\bibfield  {title} {\bibinfo {title} {{Numerical
  Representation of Planetary Ephemerides}},\ }\href@noop {} {\bibfield
  {journal} {\bibinfo  {journal} {Celestial Mechanics}\ }\textbf {\bibinfo
  {volume} {45}},\ \bibinfo {pages} {305} (\bibinfo {year} {1989})}\BibitemShut
  {NoStop}%
\bibitem [{\citenamefont {{Li}}\ and\ \citenamefont
  {Tian}(2004)}]{Li&Tian:2004}%
  \BibitemOpen
  \bibfield  {author} {\bibinfo {author} {\bibfnamefont {G.}~\bibnamefont
  {{Li}}}\ and\ \bibinfo {author} {\bibfnamefont {L.}~\bibnamefont {Tian}},\
  }\bibfield  {title} {\bibinfo {title} {{PMOE 2003 Planetary ephemeris
  framework (V) creating and using of ephemeris files}},\ }\href@noop {}
  {\bibfield  {journal} {\bibinfo  {journal} {Publication of Purple Mountain
  Astronomical Observatory}\ }\textbf {\bibinfo {volume} {23}},\ \bibinfo
  {pages} {160} (\bibinfo {year} {2004})}\BibitemShut {NoStop}%
\bibitem [{\citenamefont {Prince}\ \emph {et~al.}(2002)\citenamefont {Prince},
  \citenamefont {Tinto}, \citenamefont {Larson},\ and\ \citenamefont
  {Armstrong}}]{Prince:2002hp}%
  \BibitemOpen
  \bibfield  {author} {\bibinfo {author} {\bibfnamefont {T.~A.}\ \bibnamefont
  {Prince}}, \bibinfo {author} {\bibfnamefont {M.}~\bibnamefont {Tinto}},
  \bibinfo {author} {\bibfnamefont {S.~L.}\ \bibnamefont {Larson}},\ and\
  \bibinfo {author} {\bibfnamefont {J.~W.}\ \bibnamefont {Armstrong}},\
  }\bibfield  {title} {\bibinfo {title} {{The LISA optimal sensitivity}},\
  }\href {https://doi.org/10.1103/PhysRevD.66.122002} {\bibfield  {journal}
  {\bibinfo  {journal} {Phys. Rev. D}\ }\textbf {\bibinfo {volume} {66}},\
  \bibinfo {pages} {122002} (\bibinfo {year} {2002})},\ \Eprint
  {https://arxiv.org/abs/gr-qc/0209039} {arXiv:gr-qc/0209039 [gr-qc]}
  \BibitemShut {NoStop}%
\bibitem [{\citenamefont {Vallisneri}\ \emph {et~al.}(2008)\citenamefont
  {Vallisneri}, \citenamefont {Crowder},\ and\ \citenamefont
  {Tinto}}]{Vallisneri:2007xa}%
  \BibitemOpen
  \bibfield  {author} {\bibinfo {author} {\bibfnamefont {M.}~\bibnamefont
  {Vallisneri}}, \bibinfo {author} {\bibfnamefont {J.}~\bibnamefont
  {Crowder}},\ and\ \bibinfo {author} {\bibfnamefont {M.}~\bibnamefont
  {Tinto}},\ }\bibfield  {title} {\bibinfo {title} {{Sensitivity and
  parameter-estimation precision for alternate LISA configurations}},\ }\href
  {https://doi.org/10.1088/0264-9381/25/6/065005} {\bibfield  {journal}
  {\bibinfo  {journal} {Class. Quant. Grav.}\ }\textbf {\bibinfo {volume}
  {25}},\ \bibinfo {pages} {065005} (\bibinfo {year} {2008})},\ \Eprint
  {https://arxiv.org/abs/0710.4369} {arXiv:0710.4369 [gr-qc]} \BibitemShut
  {NoStop}%
\bibitem [{\citenamefont {{Estabrook}}\ and\ \citenamefont
  {{Wahlquist}}(1975)}]{1975GReGr...6..439E}%
  \BibitemOpen
  \bibfield  {author} {\bibinfo {author} {\bibfnamefont {F.~B.}\ \bibnamefont
  {{Estabrook}}}\ and\ \bibinfo {author} {\bibfnamefont {H.~D.}\ \bibnamefont
  {{Wahlquist}}},\ }\bibfield  {title} {\bibinfo {title} {{Response of Doppler
  spacecraft tracking to gravitational radiation.}},\ }\href
  {https://doi.org/10.1007/BF00762449} {\bibfield  {journal} {\bibinfo
  {journal} {General Relativity and Gravitation}\ }\textbf {\bibinfo {volume}
  {6}},\ \bibinfo {pages} {439} (\bibinfo {year} {1975})}\BibitemShut {NoStop}%
\bibitem [{\citenamefont {{Wahlquist}}(1987)}]{1987GReGr..19.1101W}%
  \BibitemOpen
  \bibfield  {author} {\bibinfo {author} {\bibfnamefont {H.}~\bibnamefont
  {{Wahlquist}}},\ }\bibfield  {title} {\bibinfo {title} {{The Doppler response
  to gravitational waves from a binary star source.}},\ }\href
  {https://doi.org/10.1007/BF00759146} {\bibfield  {journal} {\bibinfo
  {journal} {General Relativity and Gravitation}\ }\textbf {\bibinfo {volume}
  {19}},\ \bibinfo {pages} {1101} (\bibinfo {year} {1987})}\BibitemShut
  {NoStop}%
\bibitem [{\citenamefont {Vallisneri}\ and\ \citenamefont
  {Galley}(2012)}]{Vallisneri:2012np}%
  \BibitemOpen
  \bibfield  {author} {\bibinfo {author} {\bibfnamefont {M.}~\bibnamefont
  {Vallisneri}}\ and\ \bibinfo {author} {\bibfnamefont {C.~R.}\ \bibnamefont
  {Galley}},\ }\bibfield  {title} {\bibinfo {title} {{Non-sky-averaged
  sensitivity curves for space-based gravitational-wave observatories}},\
  }\href {https://doi.org/10.1088/0264-9381/29/12/124015} {\bibfield  {journal}
  {\bibinfo  {journal} {Class. Quant. Grav.}\ }\textbf {\bibinfo {volume}
  {29}},\ \bibinfo {pages} {124015} (\bibinfo {year} {2012})},\ \Eprint
  {https://arxiv.org/abs/1201.3684} {arXiv:1201.3684 [gr-qc]} \BibitemShut
  {NoStop}%
\end{thebibliography}%

\end{document}